\documentclass[fleqn,usenatbib]{mnras}

\usepackage{newtxtext,newtxmath}

\usepackage[T1]{fontenc}

\DeclareRobustCommand{\VAN}[3]{#2}
\let\VANthebibliography\thebibliography
\def\thebibliography{\DeclareRobustCommand{\VAN}[3]{##3}\VANthebibliography}

\usepackage{graphicx}	
\usepackage{amsmath}	
\usepackage{amssymb}	
\usepackage{tablefootnote}  

\usepackage{threeparttable} 

\title[CN and HCN intensity ratios]{Observed CN and HCN intensity ratios exhibit subtle variations in extreme galaxy environments}

\author[Blake Ledger et al.]{
B. Ledger,$^{1}$\thanks{E-mail: ledgeb1@mcmaster.}
C. D. Wilson,$^{1}$
T. Michiyama,$^{2}$
D. Iono,$^{3,4}$
S. Aalto,$^{5}$
T. Saito,$^{6}$
A. Bemis,$^{7}$
R. Aladro$^{8}$
\\
$^{1}$Department of Physics and Astronomy, McMaster University, 1280 Main St. W., Hamilton, Ontario L8S 4M1, Canada\\
$^{2}$Kavli Institute for Astronomy and Astrophysics, Peking University, 5 Yiheyuan Road, Haidian District, Beijing 100871, P.R.China\\
$^{3}$National Astronomical Observatory of Japan, National Institutes of Natural Sciences, 2-21-1 Osawa, Mitaka, Tokyo, 181-8588\\
$^{4}$Department of Astronomical Science, The Graduate University for Advanced Studies, SOKENDAI, 2-21-1 Osawa, Mitaka, Tokyo
181-8588\\
$^{5}$Department of Space, Earth and Environment, Chalmers University of Technology, Onsala Observatory, SE-439 92 Onsala, Sweden\\
$^{6}$Max-Planck-Institut f\"ur Astronomie, K\"onigstuhl 17, 69117 Heidelberg, Germany\\
$^{7}$Leiden Observatory, Leiden University, PO Box 9513, 2300 RA Leiden, The Netherlands\\
$^{8}$Max-Planck-Institut fur Radioastronomie, Auf dem H\"ugel 69, 53121 Bonn, Germany
}

\date{Accepted XXX. Received YYY; in original form ZZZ}

\pubyear{2021}

\begin{document}
\label{firstpage}
\pagerange{\pageref{firstpage}--\pageref{lastpage}}
\maketitle

\begin{abstract}
We use both new and archival ALMA data of three energy lines each of CN and HCN to explore intensity ratios in dense gas in NGC 3256, NGC 7469, and IRAS 13120-5453. The HCN (3-2)/HCN (1-0) intensity ratio varies in NGC 3256 and NGC 7469, with superlinear trends of $1.53\pm0.07$ and $1.55\pm0.05$, respectively. We find an offset to higher HCN (3-2)/HCN (1-0) intensity ratios ($\sim 0.8$) in IRAS 13120-5453 compared to NGC 3256 ($\sim 0.3-0.4$) and NGC 7469 ($\sim 0.3-0.5$). The HCN (4-3)/HCN (3-2) intensity ratio in NGC 7469 has a slope of $1.34\pm0.05$. We attribute the variation within NGC 3256 to excitation associated with the northern and southern nuclei. In NGC 7469, the variations are localized to the region surrounding the active galactic nucleus. At our resolution ($\sim700$ pc), IRAS 13120-5453 shows little variation in the HCN intensity ratios. Individual galaxies show nearly constant CN (2-1)/CN (1-0) intensity ratios. We find an offset to lower CN (2-1)/CN (1-0) intensity ratios ($\sim0.5$) in NGC 3256 compared to the other two galaxies ($\sim0.8$). For the CN (3-2)/CN (2-1) intensity ratio, NGC 7469 has a superlinear trend of $1.55\pm0.04$, with the peak localized toward the active galactic nucleus. We find high ($\sim1.7$) CN (1-0)/HCN (1-0) intensity ratios in IRAS 13120-5453 and in the northern nucleus of NGC 3256, compared to a more constant ratio ($\sim1.1$) in NGC 7469 and non-starbursting regions of NGC 3256. 
\end{abstract}




\section{Introduction}
\label{sec:introduction}
In dense regions of molecular clouds, observers use molecules with high critical densities such as HCN, HCO$^+$, and CS to identify regions of active and future star formation \citep{Wu2010, Kennicutt2012}. In particular, HCN luminosity has been shown to correlate well with total far-infrared (FIR) luminosity, a tracer of the star formation rate in galaxies \citep{Gao2004, Wu2005}, especially in intense starbursts and Ultra-Luminous and Luminous Infrared Galaxies (U/LIRGs). Sub-mm thermal pumping of HCN has also been shown to increase its excitation in the presence of strong X-ray emission around active galactic nuclei (AGN) in these luminous galaxies \citep{Kohno2005, Izumi2013, Izumi2016}.

An interesting molecule that is astrochemically related to HCN is the cyanide radical (CN). CN also has a high critical density and is primarily formed from photodissociation of HCN and neutral-neutral reactions with N, C$_{2}$, CH$_{2}$, and CH \citep{Aalto2002, Boger2005, Chapillon2012}. Intermediate stages in the reaction pathways involve neutral and ionized carbon (C and C$^{+}$; \citealt{Boger2005}).  CN is thus thought to preferentially form in regions illuminated by intense radiation fields, including the ultra-violet (UV) radiation fields of photo-dissociation regions (PDRs) surrounding massive stars \citep{Fuente1993, Fuente1995, Greaves1996, Bachiller1997, Rodriguez1998, Boger2005, Ginard2015}. CN will also be more likely to form when exposed to radiation in X-ray dominated regions (XDRs) near AGN \citep{Meijerink2005, Meijerink2007, Garcia2010} and increased cosmic ray ionization rates \citep{Boger2005, Bayet2011, Aladro2013}. The CN/HCN abundance ratio is therefore predicted to increase where photodestruction rates of HCN are high and carbon is ionized, and to decrease with cloud depth as HCN is protected from external radiation \citep{Sternberg1995, Rodriguez1998, Boger2005}.

Galaxies with starbursts and/or AGN are prime environments for probing CN and HCN and their interactions with radiation fields. U/LIRGs are typically interacting or merging galaxies characterized by high infrared luminosities \citep{Sanders2003}, large fractions of dense molecular gas \citep{Solomon1992, Gao2004, Privon2017, Sliwa2017}, and high star formation rate surface densities ($\Sigma_{\text{SFR}}$, \citealt{Vollmer2017, Privon2017}). Additionally, many of these systems host AGN, indicating the presence of both PDRs and XDRs. We refer the reader to \citet{Lonsdale2006} and \citet{Perez2020} for comprehensive reviews on U/LIRGs. U/LIRGs are typically found at large distances, but the high sensitivity and spatial resolution of modern telescopes, like the Atacama Large Millimeter/Sub-millimeter Array (ALMA), have made these galaxies accessible for detailed observations. With ALMA, we can achieve spatial resolutions that allow us to identify and isolate starburst regions and AGN in these extreme galaxy environments.

In this paper, we present the results of an observational study of the intensity ratios of CN and HCN in three U/LIRGs. We use both new and archival ALMA observations of three lines each of CN and HCN. The higher energy lines allow us to explore the interaction between these molecules in a variety of radiation environments. Section \ref{sec:observations} describes our galaxy sample, observations, data reduction, imaging, and analysis. Section \ref{sec:ratio_resuls} presents the observed intensity ratios of CN and HCN, the line ratio maps, and scatter plots of the line intensities. We briefly discuss potential physical drivers of the ratios in Section \ref{sec:discussion}, with a summary of our conclusions in Section \ref{sec:conclusions}. We will present non-LTE radiative transfer modelling directly comparing abundance ratios to PDR and XDR models, such as those from \citet{Boger2005}, \citet{Meijerink2005}, and \citet{Meijerink2007}, in a follow-up paper. Throughout this paper, our discussion of ``ratios'' refers to ``integrated line intensity ratios''and not ``molecular abundance ratios'', unless otherwise indicated.

\begin{table}
\centering
    \caption{\small{Basic properties of the galaxy sample.}}
    \begin{tabular}{cccc}
    \hline 
        \small{Property\textsuperscript{\textit{a}}} & \small{NGC 3256} & \small{NGC 7469} & \small{IRAS 13120} \\
        \hline
        Luminosity class & LIRG & LIRG & ULIRG \\
        log($L_\textsubscript{IR}$) (L$_\odot$)\textsuperscript{\textit{b}} & $11.75$ & $11.60$ & $12.29$  \\ 
        AGN contribution\textsuperscript{\textit{c}} & $<5$\% & $32-40$\% & $17-33$\% \\
        RA (J2000) & 10h27m51.3s & 23h03m15.6s & 13h15m06.4s \\
        DEC (J2000) & $-43^{\circ}54'13.5''$ & $+08^{\circ}52'26''$ & $-55^{\circ}09'22.6''$ \\
        Diameter & $3.8'\times2.1'$ & $1.5'\times1.1'$ & $0.33'\times0.33'$ \\
        Redshift & $0.00935$ & $0.01632$ & $0.03076$ \\
        $D_{\textsubscript{L}}$ (Mpc)\textsuperscript{\textit{d}} & $44$ & $66$ & $134$ \\
        $\langle$SFR$\rangle$ (M$_\odot$ yr${^{-1}}$)\textsuperscript{\textit{e}} & $84$ & $60$ & $292$ \\
    \hline
    \end{tabular}
    \begin{tablenotes}
        \item \textit{Notes:} \textsuperscript{\textit{a}}NGC 3256 and NGC 7469 properties are retrieved from the NASA/IPAC Extragalactic Database (NED\tablefootnote{http://ned.ipac.caltech.edu/}). IRAS 13120 properties come from Simbad\tablefootnote{http://simbad.u-strasbg.fr/simbad/}.
        \item \textsuperscript{\textit{b}}$L_{\textsubscript{IR}}$ data from \citet{Sanders2003} and are corrected for luminosity distance.
        \item \textsuperscript{\textit{c}}The AGN contribution to $L_\text{IR}$ is estimated from 6 \textmu m and 24 \textmu m emission for NGC 3256 and NGC 7469 \citep{Alonso2012}. For IRAS 13120, the AGN contribution is estimated from 15 \textmu m and 30 \textmu m \citep{Veilleux2013}, 60 \textmu m \citep{Teng2015}, and total 8-1000 \textmu m emission \citep{Iwasawa2011}.
        \item \textsuperscript{\textit{d}}Luminosity distances from redshifts (corrected to the 3K CMB reference frame) and assuming H$_0$ = 70.5 km s$^{-1}$ Mpc$^{-1}$. For NGC 7469, the SN Type Ia distance is from \citet{Ganeshalingam2013}.
        \item \textsuperscript{\textit{e}}SFRs calculated from $L_{\text{IR}}$ using Equation 12 in \citet{Kennicutt2012}; not corrected for AGN contribution.
    \end{tablenotes}
    \label{tab:galaxy_props}
\end{table}

\section{Observations and Data Reduction}
\label{sec:observations}

\subsection{U/LIRG sample}
\label{sec:sample}
Our sample consists of two LIRGs, NGC 3256 and NGC 7469, and one ULIRG, IRAS 13120-5453 (hereafter referred to as IRAS 13120). The individual properties of these galaxies are summarized in Table \ref{tab:galaxy_props}. The different galaxy environments available in this sample allow us to probe the effects of both UV and X-ray radiation fields on the molecular gas.

NGC 3256 is a nearby luminous galaxy merger that can be separated into northern and southern nuclear regions. The northern nucleus is fairly face-on and has a nuclear disk with significant starburst activity \citep{Sakamoto2014}. The southern nucleus is mostly edge-on and thought to host an embedded dormant AGN \citep{Sakamoto2014}. NGC 3256 is a complicated system containing molecular outflows from both nuclei \citep{Sakamoto2014} and high fractions of shocked and dense gas \citep{Harada2018}. \citet{Sakamoto2014} find that the outflows in the southern nucleus are highly collimated, bipolar nuclear jets with velocities of $\sim2000$ km s$^{-1}$. \citet{Brunetti2021} argue that, due to a lack of significant trends in molecular gas surface density, brightness temperature, and velocity dispersion with physical scale, NGC 3256 must contain a smooth interstellar medium down to $\sim55$ pc scales.

NGC 7469 hosts a nuclear luminous type-1 AGN ($L_{\textsubscript{2-10keV}} = 1.5\times10^{43}$ erg s$^{-1}$; \citealt{Liu2014}) and has a starburst ring surrounding the central portion of the galaxy \citep{Izumi2015, Izumi2020}. The AGN is creating an XDR $\sim50$ pc in size in the nuclear region, and this XDR is surrounded by a circumnuclear disk of star-forming gas \citep{Izumi2020}. \citet{Izumi2015} find the integrated intensities of HCN (4-3), HCO$^{+}$ (4-3), and CS (7-6) are higher in the nucleus than in the starburst ring and predict that the HCN abundance is enhanced in the nucleus due to increased sub-mm radiation near the AGN \citep{Izumi2013, Izumi2016}.

IRAS 13120 has been optically classified as a Seyfert 2 galaxy \citep{Veron2010}. \citet{Teng2015} find that observations of IRAS 13120 are consistent with the galaxy hosting an inactive, Compton-thick AGN ($N_{\text{H}} > 10^{24}$ cm$^{-2}$). $^{12}$CO observations by \citet{Sliwa2017} show evidence for a young starburst ($<7$ Myr) in the central $\sim500$ pc region. \citet{Privon2017} find the HCN/HCO$^{+}$ ratio is higher in the central starburst and suggest this is from mechanical heating of the gas by supernovae feedback.

\subsection{Data and imaging}
\label{sec:data_and_imaging}

\tabcolsep=0.13cm
\begin{table}
\centering
    \caption{\small{New and archival ALMA projects with HCN (J = 1-0, 3-2, 4-3) and CN (N = 1-0, 2-1, 3-2) observations of the three target galaxies.}}
    \begin{tabular}{cccc}
    \hline
        \small{Line} & \small{Project code} & \small{PI} & \small{ALMA data reference} \\
        \hline
        NGC 3256 & & & \\
        \hline
        HCN (1-0) & 2015.1.00993.S & Michiyama, T. & \citet{Michiyama2018} \\
        HCN (3-2) & 2015.1.00412.S & Harada, N. & \citet{Harada2018} \\
        HCN (4-3) & 2018.1.00493.S & Wilson, C. & This paper \\
        CN (1-0) & 2011.0.00525.S & Sakamoto, K. & \citet{Sakamoto2014} \\
        CN (2-1) & 2015.1.00412.S & Harada, N. & \citet{Harada2018} \\
        CN (3-2) & 2018.1.00493.S & Wilson, C. & This paper \\
        \hline
        NGC 7469 & & & \\
        \hline
        HCN (1-0) & 2012.1.00165.S & Izumi T. & \citet{Izumi2015} \\
        HCN (3-2) & 2012.1.00034.S & Imanishi M. & \citet{Imanishi2016} \\
        HCN (4-3) & 2012.1.00165.S & Izumi T. & \citet{Izumi2015} \\
        CN (1-0) & 2013.1.00218.S & Izumi T. & \citet{Wilson2019}\textsuperscript{\textit{a}} \\
        CN (2-1) & 2015.1.00412.S & Harada, N. & \citet{Harada2018} \\
        CN (3-2) & 2018.1.00493.S & Wilson, C. & This paper \\
        \hline
        IRAS 13120 & & & \\
        \hline
        HCN (1-0) & 2013.1.00379.S & Sliwa, K. & \citet{Sliwa2017} \\
        HCN (3-2) & 2018.1.00493.S & Wilson, C. & This paper \\
        HCN (4-3) & 2015.1.00102.S & Iono D. & \citet{Fluetsch2019}\textsuperscript{\textit{b}} \\
        CN (1-0) & 2015.1.00287.S & Sliwa, K. & \citet{Wilson2019}\textsuperscript{\textit{a}} \\
        CN (2-1) & 2016.1.00777.S & Sliwa, K. & $-$ \\
        CN (3-2) & 2018.1.00493.S & Wilson, C. & This paper \\
    \hline
    \end{tabular}
    \label{tab:Project_IDs}
    \begin{tablenotes}
        \small
            \item \textit{Notes:} \textsuperscript{\textit{a}}Reference uses the CO (1-0) transition from these project IDs. 
            \item \textsuperscript{\textit{b}}Reference uses the CO (3-2) transition from this project ID.
    \end{tablenotes}
\end{table}

Table \ref{tab:Project_IDs} lists the project IDs for the new and archival ALMA data used in our analysis. The data were reduced and imaged with the Common Astronomy Software Application (CASA; \citealt{McMullin2007}). The raw \textit{uv} data were calibrated for each project using the relevant CASA version. All subsequent data reduction was performed using CASA version 5.6.1. Continuum subtraction was performed on each \textit{uv} dataset using line-free channels and CASA's \textsc{uvcontsub} task.

Given the large collection of ALMA observations from which our data were obtained, we put significant effort into matching the spatial and spectral resolutions of the lines. The \textit{uv} data for each galaxy were imaged individually with CASA's \textsc{tclean} task. All imaging was done using Brigg's weighting \citep{Briggs1995}. Making use of \textsc{tlcean}'s \textsc{cell}, \textsc{uvtaper}, \textsc{uvrange}, and \textsc{width} parameters, the \textit{uv} coverage, spectral channel width, and dirty beam sizes were compared and matched for each galaxy. The limiting factors were the largest minimum \textit{uv} coverage, the largest spectral channel width, and the largest dirty beam size. We note that applying a \textit{uv} cut-off and taper with the \textsc{uvrange} and \textsc{uvtaper} parameters in \textsc{tclean} will limit the number of short baselines in our data and thus could potentially lead to missing flux on the largest angular scales. The \textsc{uvrange} parameter was matched between all lines in each galaxy in order to recover similar flux scales for each line. We tried to limit the problem of missing flux in our interferometric observations by choosing the largest minimum \textsc{uvrange} cut-off while still matching all lines.

A measure of the mean RMS noise was found from both the dirty and final smoothed image cubes for each spectral line in CASA using the \textsc{imstat} task on line-free channels. The sensitivities varied for each line in each galaxy depending on the specific ALMA observations and can be found in Tables \ref{tab:imaging_props} and \ref{tab:imaging_vel_props}. The imaging of all galaxies was limited to regions where the primary beam response was greater than 20\% and cleaning was performed using CASA's \textsc{auto-multithresh} \citep{Kepley2019} down to the 2$\sigma$ level using the line sensitivities from the dirty cubes. Masks produced by \textsc{tclean}'s auto-masking algorithm were checked for consistency during major cleaning cycles. Most observations were completed with a single pointing and imaged with matched phase centres in \textsc{tclean}, except for the ALMA Band 7 observations of CN (3-2) and HCN (4-3) in NGC 3256 which had multiple pointings. In these cases, the imaging was done as a mosaic with the same matched phase centres. A final smoothing to a common round beam was completed for all lines in each galaxy with the CASA task \textsc{imsmooth}.

\tabcolsep=0.13cm
\begin{table}
\centering
    \caption{\small{Data reduction imaging properties and line sensitivities.}}
    \begin{tabular}{cccc}
    \hline 
        \small{Imaging property} & \small{NGC 3256} & \small{NGC 7469} & \small{IRAS 13120} \\
        \hline
        Beam size\textsuperscript{\textit{a}} & $2.2''$ & $0.95''$ & $1.1''$ \\
        Beam (pc) & $469$ & $304$ & $715$ \\
        Velocity resolution (km s$^{-1}$)\textsuperscript{\textit{b}} & $26.43$ & $20.68$ & $20.64$ \\
        Pixel size\textsuperscript{\textit{b}} & $0.3''$ & $0.15''$ & $0.15''$ \\
        Re-binned pixel size\textsuperscript{\textit{b}} & $1.1''$ & $0.475''$ & $0.55''$ \\
        \textsc{uvrange} cut-off (k$\lambda$)\textsuperscript{\textit{c}} & $>15$ & $>19.8$ & $>15.9$ \\
        Maximum recoverable scale\textsuperscript{\textit{c}} & $8.25''$ & $6.25''$ & $7.78''$ \\
        Sensitivities (mJy beam$^{-1}$)\textsuperscript{\textit{d}} & & & \\
        HCN(1-0) & $0.24$ & $0.27$ & $0.95$ \\
        HCN(3-2) & $1.53$ & $0.44$ & $0.58$\\
        HCN(4-3) & $0.99$ & $0.81$ & $2.00$ \\
        CN(1-0) & $0.66$ & $0.49$ & $1.11$ \\
        CN(2-1) & $0.95$ & $0.26$ & $0.39$ \\
        CN(3-2) & $0.53$ & $0.71$ & $0.97$ \\
    \hline
    \end{tabular}
    \begin{tablenotes}
    \small
        \item \textit{Notes:} \textsuperscript{\textit{a}}Beams smoothed, rounded, and matched to this resolution for all lines.
        \item \textsuperscript{\textit{b}}The velocity resolution and pixel size were matched between all lines.
        \item \textsuperscript{\textit{c}}\textsc{uvrange} cut-off based on the minimum \textit{uv} range covered by all lines in each galaxy and has also been converted to a maximum recoverable scale of emission.
        \item \textsuperscript{\textit{d}}Sensitivities determined using line-free channels of dirty image cubes.
    \end{tablenotes}
    \label{tab:imaging_props}
\end{table}

\subsubsection{NGC 3256 processing}
\label{sec:3256_data}
The spectral resolution in NGC 3256 was limited by the CN (1-0) line at $26.43$ km s$^{-1}$ and all lines were binned in velocity to this channel width. The limiting dirty beam size was set by the HCN (1-0) observations at $2.1''\times1.7''$, PA = $88^\circ$. The parameter \textsc{uvtaper} was used to taper the longest baselines in each dataset until the beams were matched relatively well, before a final smoothing to a $2.2''$ round beam was applied. For the HCN (3-2), HCN (4-3), CN (2-1), and CN (3-2) lines, natural weighting (Brigg's with robust = 2.0) was used to increase sensitivity and naturally increase the dirty beam size before tapering. For HCN (1-0) and CN (1-0), the resolution was close enough to the target size of $2.2''$ that little tapering was required and Brigg's weighting with robust = 0.5 was sufficient.

\subsubsection{NGC 7469 processing}
\label{sec:7469_data}
The three CN transitions in NGC 7469 had limiting spectral resolutions of $\sim5.17$ km s$^{-1}$; however, the channel widths were increased to an integer multiple of $4\times5.17=20.68$ km s$^{-1}$ to smooth the data spectrally and increase the signal-to-noise. This smoothing helped match the spectral resolution of the NGC 3256 and IRAS 13120 data. The CN (1-0) line limited our resolution to a beam size of $0.89''\times0.55''$, PA = $-47.3^\circ$. We thus targeted a smoothed 0.95'' round beam, and adjusted weighting and \textsc{uvtaper} parameters to match this target. HCN (1-0), CN (1-0), and CN (3-2) were imaged using robust = 0.5. HCN (3-2), HCN (4-3), and CN (3-2) were imaged using natural weighting.

\subsubsection{IRAS 13120-5453 processing}
\label{sec:13120_data}
In the IRAS 13120 data, the native spectral resolution varied for each of the three CN lines. The HCN lines all had the same $\sim3.3$ km s$^{-1}$ velocity resolution. Channel widths were fixed at $20.64$ km s$^{-1}$, an integer multiple of the CN (1-0) transition line. This spectral smoothing enhanced the signal-to-noise ratio for the large velocity dispersions seen in IRAS 13120. The limiting beam size was the HCN (1-0) line at $1.01''\times0.51''$, PA = $73.7^\circ$, imaged with robust = 0.5. The remaining lines were imaged with robust = 2.0, as they all had dirty beams with better than $0.6''$ resolution. Tapers were applied and we smoothed the data to a final target $1.1''$ round beam.

\subsection{Integrated intensities and measured ratios}
\label{sec:integrated_intensities}

\tabcolsep=0.08cm
\begin{table}
\centering
    \caption{\small{CN and HCN measured intensity ratios.}}
    \begin{tabular}{ccccccc}
    \hline
        \small{Region} & \small{$\frac{\text{HCN (3-2)}}{\text{HCN (1-0)}}$} & \small{$\frac{\text{HCN (4-3)}}{\text{HCN (3-2)}}$} & \small{$\frac{\text{CN (2-1)}}{\text{CN (1-0)}}$} & \small{$\frac{\text{CN (3-2)}}{\text{CN (2-1)}}$} & \small{$\frac{\text{CN (1-0)}}{\text{HCN (1-0)}}$} & \small{$\frac{\text{CN (3-2)}}{\text{HCN (3-2)}}$} \\
        \hline
        \small{NGC 3256} & & & & & & \\
        \hline
        \small{N. nucl}. & $0.39(4)$ & $0.48(7)$ & $0.47(5)$ & $0.27(4)$ & $1.7(1)$ & $0.55(8)$ \\
        \small{S. nucl.} & $0.33(4)$ & $0.46(7)$ & $0.44(5)$ & $0.22(3)$ & $1.12(9)$ & $0.34(5)$ \\
        \small{Non-nucl.} & $0.27(3)$ & $0.45(6)$ & $0.43(5)$ & $0.21(3)$ & $1.14(8)$ & $0.38(5)$ \\
        \hline
        \small{NGC 7469} & & & & & & \\
        \hline
        \small{Nucl.} & $0.52(6)$ & $0.48(7)$ & $0.80(9)$ & $0.45(6)$ & $1.13(9)$ & $0.8(1)$ \\
        \small{Non-nucl.} & $0.33(4)$ & $0.37(5)$ & $0.76(9)$ & $0.30(4)$ & $1.11(8)$ & $0.8(1)$ \\
        \hline
        \small{IRAS 13120} & & & & & & \\
        \hline
        \small{Global} & $0.80(9)$ & $0.8(1)$ & $0.77(9)$ & $0.53(8)$ & $1.7(1)$ & $0.8(1)$ \\
        \small{Peak} & $0.79(9)$ & $0.9(1)$ & $0.80(9)$ & $0.57(8)$ & $1.5(1)$ & $0.8(1)$ \\
    \hline
    \end{tabular}
    \begin{tablenotes}
    \small
        \item \textit{Notes:} Uncertainties are measurement plus calibration uncertainties and are given as the uncertainty on the last digit (i.e. $0.39(4) = 0.39\pm0.04$).
        \item ALMA calibration uncertainties are 5\% (Band 3) and 10\% (Band 6 and 7).
    \end{tablenotes}
    \label{tab:measured_ratios}
\end{table}

We produced integrated intensity (moment 0) maps of all lines in our three galaxies. Cleaned, smoothed data cubes were extracted from CASA and subsequent processing, including the creation of the moment 0 maps, was done using the \textsc{astropy}\footnote{http://www.astropy.org} software in \textsc{Python} 3.7. The cubes were trimmed to include only channels with line emission (found using the CASA viewer). The HCN (1-0), HCN (3-2), and HCN (4-3) line cubes were trimmed to include the same range in velocity. The CN lines were trimmed depending on their hyperfine structure so that all hyperfine lines were included in the calculation of the integrated intensities. Exact velocity ranges for all six lines can be found in Table \ref{tab:imaging_vel_props}. An RMS cut-off of $3.5\sigma$ was used to mask the cleaned cubes to include strong emission while avoiding noisy emission in lines with lower S/N. This $3.5\sigma$ limit used the line sensitivities calculated in the smoothed cubes and was chosen to overcome the varying sensitivities between different lines in each galaxy. The moment 0 maps were further masked to only include pixels with emission in all six spectral lines. In cases where multiple hyperfine lines are present (the HCN (1-0), CN (1-0), CN (2-1), and CN (3-2) lines), the integrated intensities include all hyperfine lines that were observed. The moment 0 maps were converted to physical units of K km s$^{-1}$ and corrected for the primary beam response. Moment 0 maps for each line can be found in the Appendix (Figures \ref{fig:3256_mom0s}, \ref{fig:7469_mom0s}, and \ref{fig:13120_mom0s}). Uncertainty maps were made to match each moment 0 map. Individual uncertainties on each pixel were calculated as $\sigma=\text{rms}\cdot \sqrt{\Delta V_{\textsubscript{chan}} \cdot V_{\textsubscript{line}}}$, where $\Delta V_{\textsubscript{chan}}$ is the velocity width of an individual channel and $V_{\textsubscript{line}}$ is the width of the line integrated in that pixel. The integrated intensity maps were used to calculate total intensities and intensity ratios. Maps of the ratios covering the high S/N regions in each galaxy are shown in Figures \ref{fig:CN_and_HCN_spatial_maps} and \ref{fig:CN_HCN_spatial_maps}.

We calculate global line ratios for each galaxy using the average line intensities in K km s$^{-1}$. Along with the global ratios, we measured line ratios in the northern and southern nuclei of NGC 3256 and the nucleus of NGC 7469. The central nuclear pixels were identified using the 93 GHz radio continuum emission peak for NGC 3256 and NGC 7469. To isolate the nuclear regions, we used an aperture centred on the nuclear pixel. The size of the aperture is equal to the full-width at half-maximum of the beam in each galaxy. Additionally, for NGC 3256 and NGC 7469 we calculated non-nuclear line ratios using all non-nuclear pixels. For IRAS 13120, we measured the line ratios at a single pixel at the 93 GHz radio continuum peak. Table \ref{tab:measured_ratios} presents the results of this analysis and the measured ratios in the different regions of each galaxy.

\subsection{Pixel-by-pixel comparisons}
\label{sec:pixel-by-pixel}

\tabcolsep=0.2cm
\begin{table}
\centering
    \caption{\small{Slopes from pixel-by-pixel comparisons.\textsuperscript{\text{a}}}}
    \begin{tabular}{ccccc}
    \hline
        \small{Ratio} & \small{NGC 3256} & \small{NGC 7469} & \small{IRAS 13120} & \small{All pixels}\\
        \hline
        \small{$\frac{\text{HCN (3-2)}}{\text{HCN (1-0)}}$} & $1.53\pm0.07$ & $1.55\pm0.05$ & $0.84\pm0.06$ & $1.46\pm0.04$\\
        \hline
        \small{$\frac{\text{HCN (4-3)}}{\text{HCN (3-2)}}$} & $0.99\pm0.04$ & $1.34\pm0.05$ & $1.3\pm0.1$ & $1.12\pm0.03$\\
        \hline
        \small{$\frac{\text{CN (2-1)}}{\text{CN (1-0)}}$} & $0.95\pm0.03$ & $0.93\pm0.03$ & $0.91\pm0.02$ & $1.07\pm0.02$\\
        \hline
        \small{$\frac{\text{CN (3-2)}}{\text{CN (2-1)}}$} & $1.26\pm0.05$ & $1.55\pm0.04$ & $1.11\pm0.04$ & $1.33\pm0.02$\\
        \hline
        \small{$\frac{\text{CN (1-0)}}{\text{HCN (1-0)}}$} & $1.49\pm0.06$ & $1.26\pm0.05$ & $0.83\pm0.05$ & $1.25\pm0.03$\\
        \hline
        \small{$\frac{\text{CN (3-2)}}{\text{HCN (3-2)}}$} & $1.06\pm0.09$ & $1.17\pm0.05$ & $0.94\pm0.06$ & $1.21\pm0.04$\\
    \hline
    \end{tabular}
    \label{tab:slopes}
    \begin{tablenotes}
        \small
            \item \textit{Notes:} \textsuperscript{\text{a}}Fit parameters determined from Linmix linear fitting of the form log(y) = m log(x) + b.
    \end{tablenotes}
\end{table}

We explored correlations between spectral lines using a pixel-by-pixel comparison of the intensities. We first re-sampled the imaged data cubes to half the full-width at half-maximum of the beam. The number of pixels sampled across the beam was reduced by a factor of 3 for all galaxies, and the new pixel sizes can be found in Table \ref{tab:imaging_props}. Following the same process as described in Section \ref{sec:integrated_intensities}, moment 0 maps with associated uncertainties were made using masked, matched pixels in the re-sampled image cubes in physical units of K km s$^{-1}$. For NGC 7469, we corrected the measured intensities in the scatter plots for an inclination angle of $45\pm5$ degrees \citep{Davies2004}. The intensities of the other two galaxies were not corrected for inclination angle\footnote{NGC 3256 is a complicated system where the northern and southern nuclei are fairly face-on and edge-on, respectively ($i_{\text{N}} \approx 30^{\circ}$ and $i_{\text{S}} \approx 80^{\circ}$; \citealt{Sakamoto2014}), making it challenging to correct for any specific angle. No inclination angle for IRAS 13120 was found in the literature. We performed the same analysis applying a conservative estimate for inclination angles of $60^\circ$ to NGC 3256 and IRAS 1320 and found that the resulting slopes and intercepts match the results in Table \ref{tab:slopes} within one standard deviation.}.

\begin{figure*}
	\includegraphics[width=\textwidth]{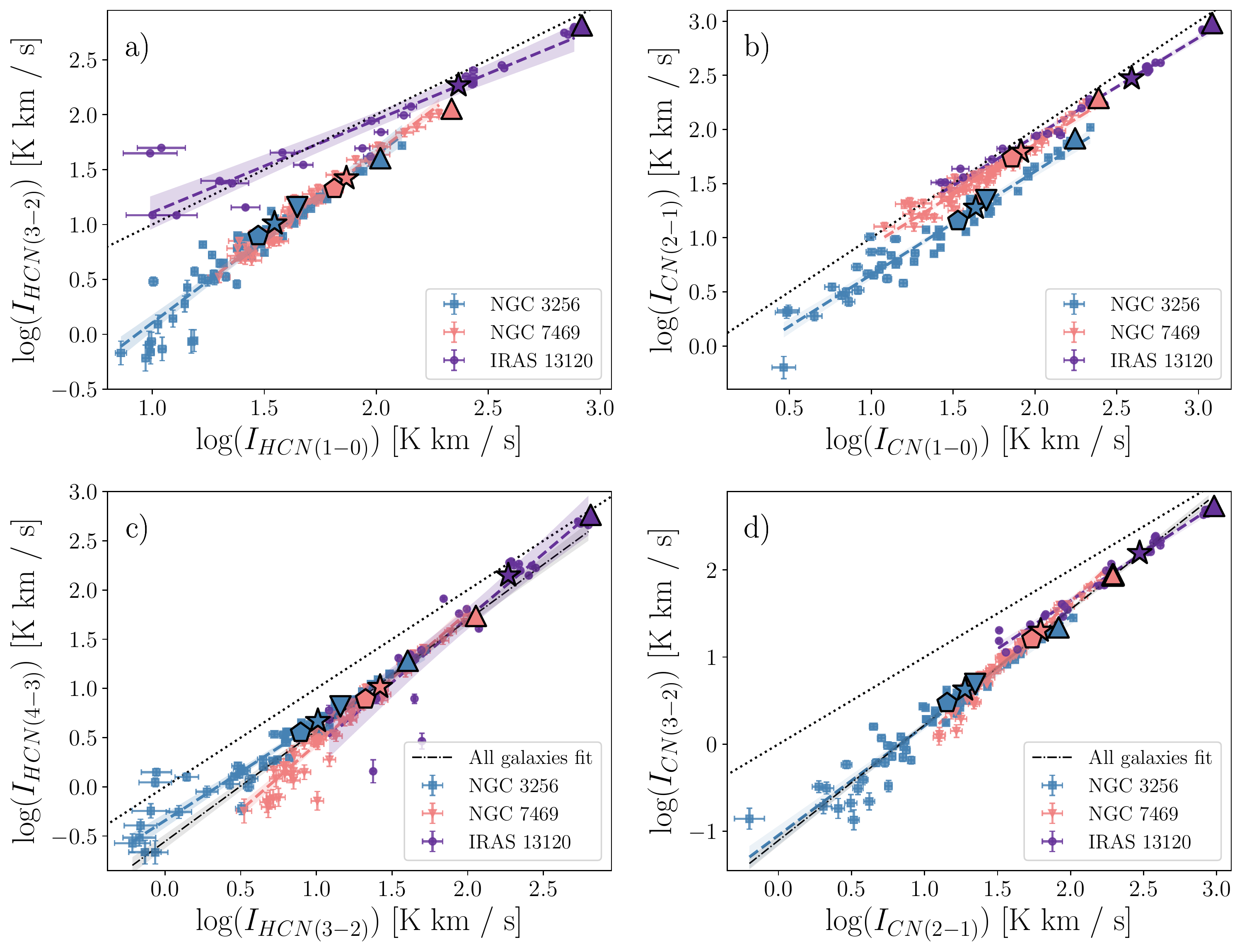}
	\caption{Pixel-by-pixel comparisons of individual line intensities in each galaxy. The higher-J/N transition of the two lines is always plotted on the y-axis. The black dotted line is the one-to-one line. Blue squares correspond to NGC 3256; pink inverted triangles correspond to NGC 7469; purple circles correspond to IRAS 13120. The dashed lines in each colour represent the Linmix fits for each galaxy, with the shaded region representing the 95\% confidence interval of these fits. Large stars show the global ratio in each galaxy and large pentagons show the non-nuclear ratios in NGC 3256 and NGC 7469. Large triangles show ratios for the northern nucleus of NGC 3256, the nucleus of NGC 7649, and the continuum peak in IRAS 13120, while the large inverted triangle shows the ratio for the southern nucleus of NGC 3256. The black dot-dashed lines and shaded regions in (c) and (d) show Linmix fits to all three galaxies. Comparison plots of the additional J/N-level transition lines of CN and HCN can be found in the Appendix.}
   \label{fig:HCN_and_CN_galax_comps}
\end{figure*}

\begin{figure*}
	\includegraphics[width=\textwidth]{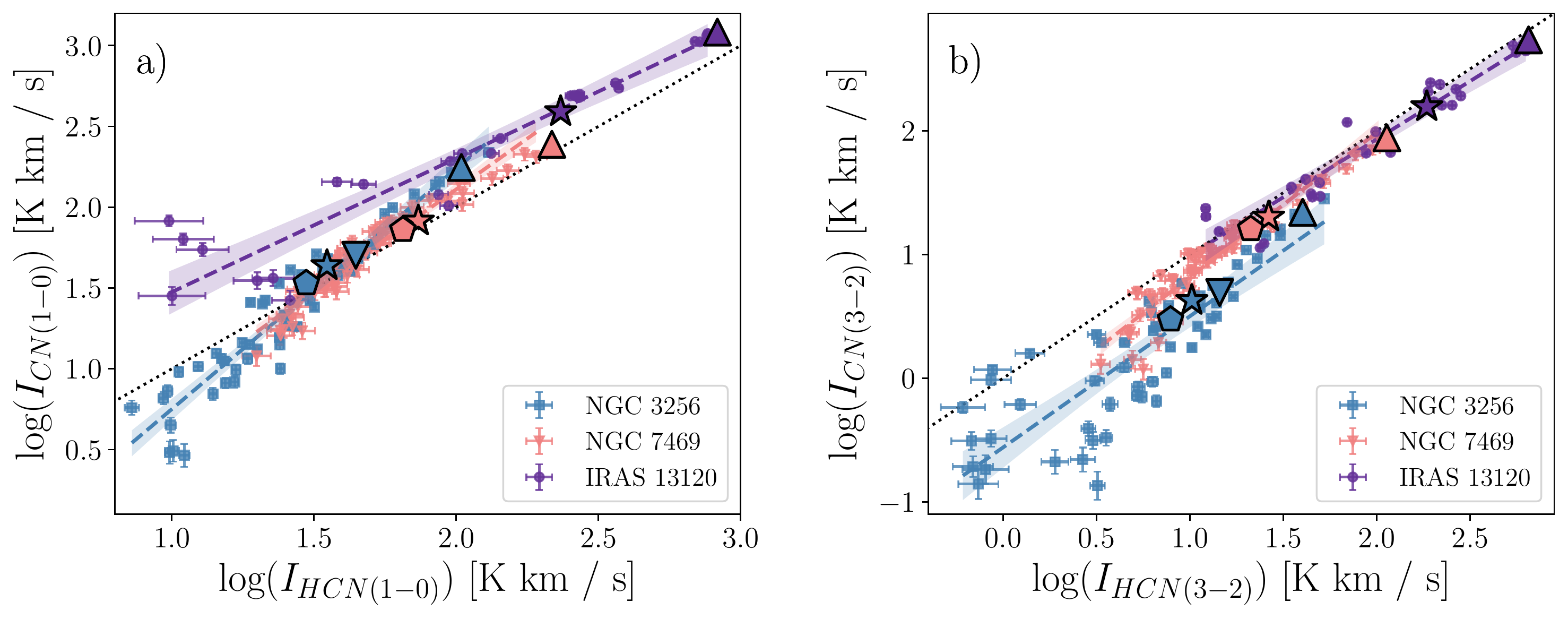}
	\caption{Pixel-by-pixel comparisons for CN compared to HCN for selected transitions. For plot descriptions, refer to the caption of Figure \ref{fig:HCN_and_CN_galax_comps}. Comparison plots of the additional J/N-level transition lines of CN and HCN can be found in the Appendix.}
   \label{fig:HCN_CN_galax_comps}
\end{figure*}

We present pixel-by-pixel comparisons of the integrated intensities in the scatter plots of Figures \ref{fig:HCN_and_CN_galax_comps} and \ref{fig:HCN_CN_galax_comps}. We used the Linmix linear regression method to fit the integrated intensity pixels in log-log space and account for the higher uncertainties seen in the lower S/N pixels \citep{Kelly2007, Meyers2018}. The resulting slopes found for each intensity ratio are presented in Table \ref{tab:slopes}. Full details of all calculated slopes and intercepts for all line ratio combinations can be found in Table \ref{tab:slopes_full} in the Appendix.

\section{Variations in the CN and HCN intensity ratios}
\label{sec:ratio_resuls}

\begin{figure*}
	\includegraphics[width=\textwidth]{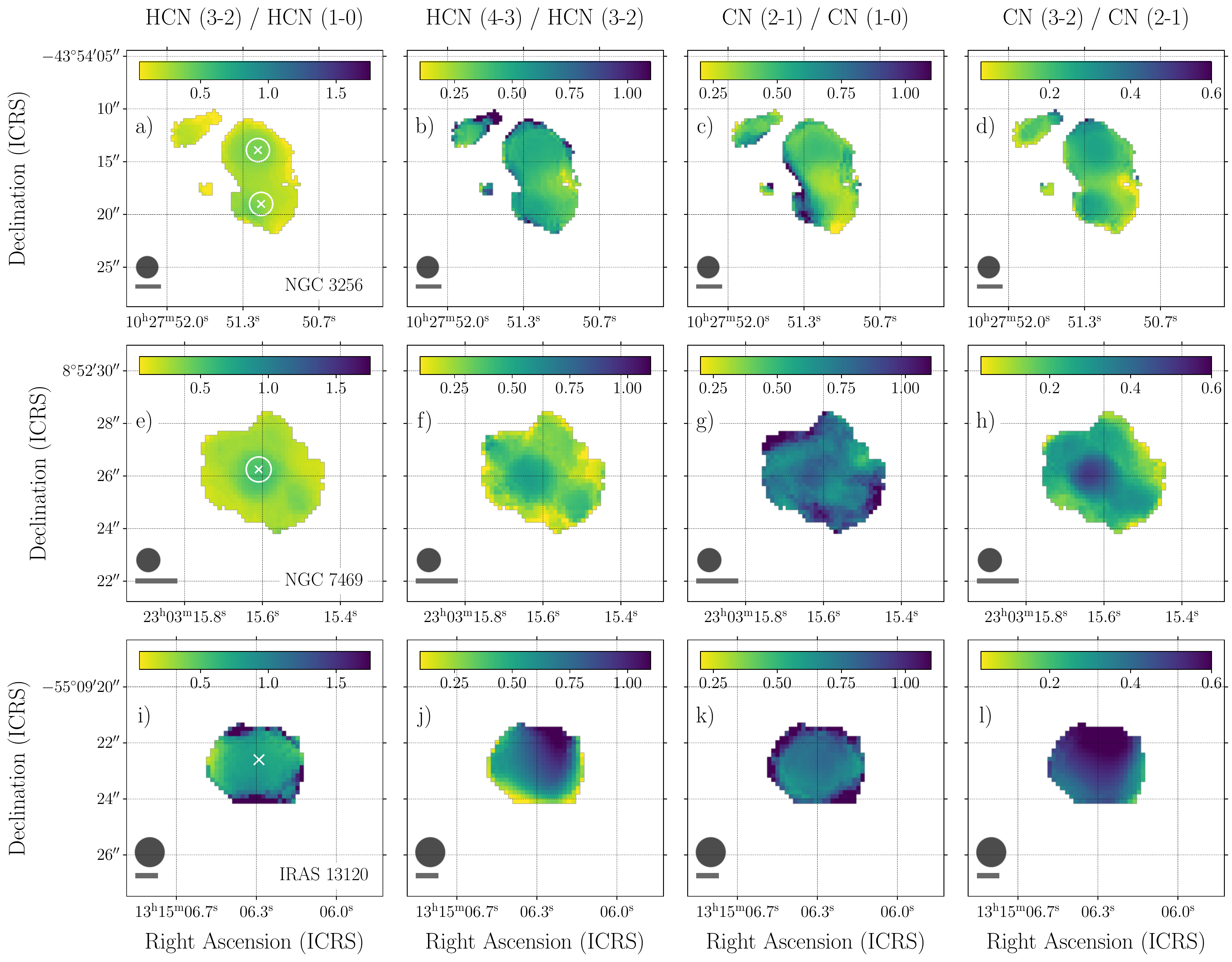}
    \caption{\textit{Top row (a-d)}: NGC 3256 ratio maps. \textit{Middle row (e-h)}: NGC 7469 ratio maps. \textit{Bottom row (i-l)}: IRAS 13120 ratio maps. Ratio maps are calculated from the integrated intensities in K km s$^{-1}$. The beam size is shown as the black circle and the scale bars are set to 500 pc. Intensity ratios in each column are (from left to right) HCN (3-2)/HCN (1-0), HCN (4-3)/HCN (3-2), CN (2-1)/CN (1-0), and CN (3-2)/CN (2-1). White crosses indicate the 93 GHz radio continuum peaks in each galaxy and white circles represent the apertures within which nuclear ratios were measured. The coordinates of the continuum peaks in each galaxy are: NGC 3256 north nucleus: (10h27m51.23s, $-43^{\circ}54'14.23''$); NGC 3256 south nucleus: (10h27m51.20s, $-43^{\circ}54'19.29''$); NGC 7469 nucleus: (23h3m15.62s, $8^{\circ}52'26.10''$); IRAS 13120 nucleus: (13h15m6.34s, $-55^{\circ}9'22.75''$).}
    \label{fig:CN_and_HCN_spatial_maps}
\end{figure*}

\begin{figure*}
	\includegraphics[width=\textwidth]{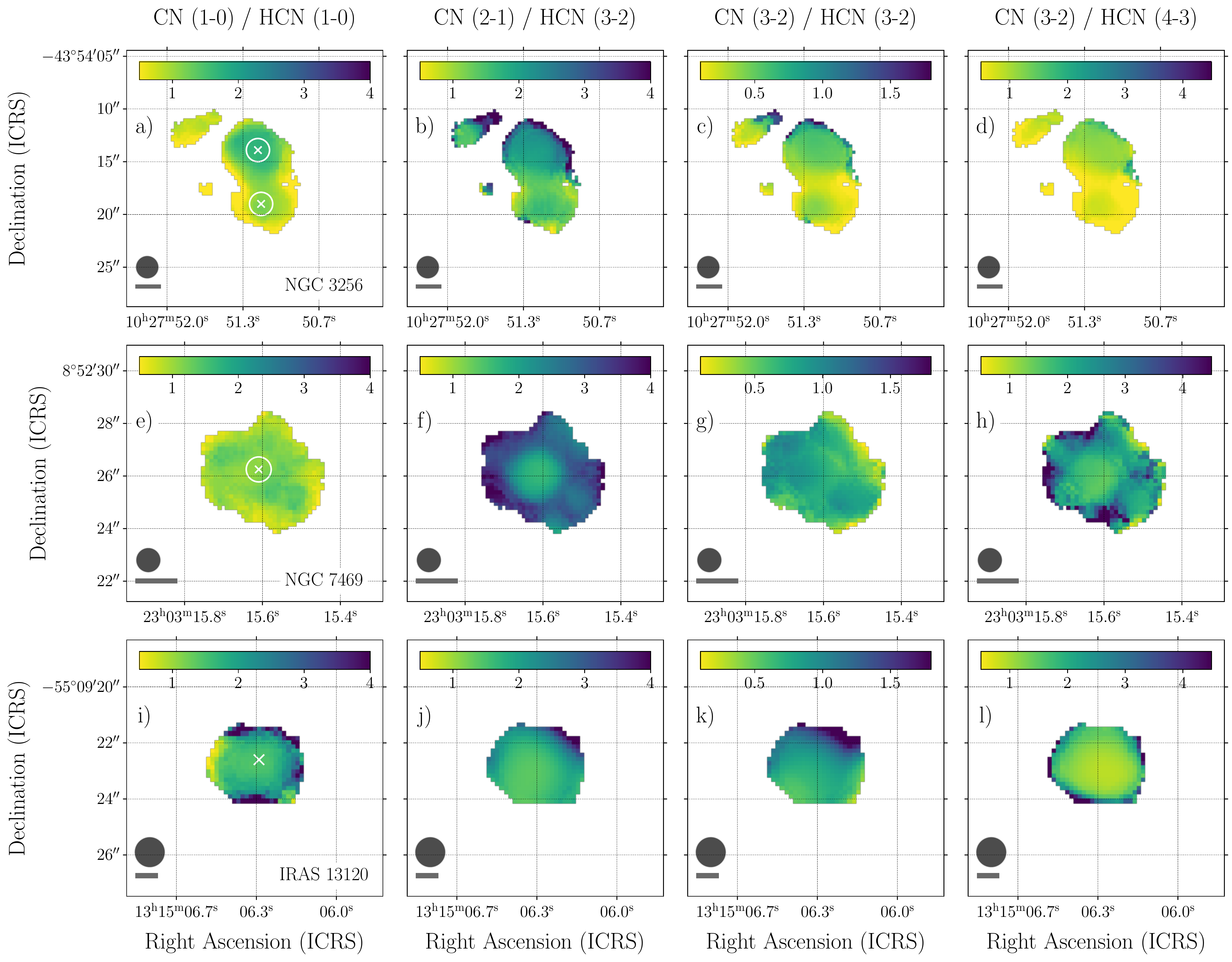}
    \caption{\textit{Top row (a-d)}: NGC 3256 ratio maps. \textit{Middle row (e-h)}: NGC 7469 ratio maps. \textit{Bottom row (i-l)}: IRAS 13120 ratio maps. Ratio maps are calculated from the integrated intensities in K km s$^{-1}$. Beam sizes, scale bars, white circles, and crosses are as described in Figure \ref{fig:CN_and_HCN_spatial_maps}. Intensity ratios in each column are (from left to right) CN (1-0)/HCN (1-0), CN (2-1)/HCN (3-2), CN (3-2)/HCN (3-2), and CN (3-2)/HCN (4-3).}
    \label{fig:CN_HCN_spatial_maps}
\end{figure*}

In this section, we present the measured intensity ratios and slopes from our CN and HCN line analysis. Discussion and physical interpretation of the ratios is deferred until Section \ref{sec:discussion}.

\subsection{Regions of enhanced HCN ratios}
\label{sec:HCN_excit_results}

We find variations in the HCN (3-2)/HCN (1-0) ratio in NGC 3256 and NGC 7469 (Figure \ref{fig:HCN_and_CN_galax_comps}a), with spatial variations in the ratio within these galaxies (Figures \ref{fig:CN_and_HCN_spatial_maps}a and \ref{fig:CN_and_HCN_spatial_maps}e). Both galaxies show a similar superlinear trend ($1.53\pm0.07$ for NGC 3256, $1.55\pm0.05$ for NGC 7469). The northern nucleus of NGC 3256 shows a marginal increase in the HCN (3-2)/HCN (1-0) ratio to a value of $0.39\pm0.04$, in contrast to the value of $0.27\pm0.03$ in the non-nuclear pixels. In NGC 7469, the HCN (3-2)/HCN (1-0) ratio is significantly higher ($0.52\pm0.06$) in the central $\sim500$ pc compared to the value of the non-nuclear pixels ($0.33\pm0.04$) in the rest of the disk.

Comparing the two galaxies, the HCN (3-2)/HCN (1-0) and HCN (4-3)/HCN (1-0) global and non-nuclear ratios in NGC 3256 and NGC 7469 are similar within our uncertainties (Table \ref{tab:measured_ratios} and \ref{tab:measured_ratios_full}). In the nuclear regions, however, both ratios are higher in NGC 7469 than in NGC 3256.

There is no difference in the HCN (3-2)/HCN (1-0) ratio for IRAS 13120 between the global ratio and the continuum peak. IRAS 13120 shows a slight sublinear trend in the HCN (3-2)/HCN (1-0) ratio, with a slope of $0.84\pm0.06$ (Figure \ref{fig:HCN_and_CN_galax_comps}a), significantly different from the superlinear trends with slopes $>1.5$ found in NGC 3256 and NGC 7469. The HCN (3-2)/HCN (1-0) ratio of the ULIRG is higher ($\sim0.8$) than both the global and nuclear values of the LIRGs ($\sim0.3-0.5$). 

The HCN (4-3)/HCN (3-2) ratio in NGC 3256 has a slope consistent with unity ($0.99\pm0.04)$. This ratio in NGC 3256 has nearly identical values of $0.48\pm0.07$, $0.46\pm0.07$, and $0.45\pm0.06$ in the northern nucleus, southern nucleus, and non-nuclear pixels, respectively. Slight variations can be seen in the HCN (4-3)/HCN (3-2) ratio in NGC 3256 in Figure \ref{fig:CN_and_HCN_spatial_maps}. In contrast, NGC 7469 and IRAS 13120 show variations in the HCN (4-3)/HCN (3-2) ratio with superlinear trends of $1.34\pm0.05$ and $1.3\pm0.1$, respectively. A fit to the pixels from all galaxies in the HCN (4-3)/HCN (3-2) ratio gives a slightly superlinear slope of $1.13\pm0.03$ (Figure \ref{fig:HCN_and_CN_galax_comps}c).

\subsection{Consistency in the CN ratios}
\label{sec:CN_sim_results}

The slopes for the CN (2-1)/CN (1-0) ratio are consistent within one standard deviation for all three galaxies in our sample, with an average value of $0.93\pm0.05$ (Figure \ref{fig:HCN_and_CN_galax_comps}b). We do not resolve any structure in the line ratio maps of individual galaxies (\ref{fig:CN_and_HCN_spatial_maps}). There is, however, an offset to lower values for the CN (2-1)/CN (1-0) intensity ratio in NGC 3256 ($\sim0.4$) compared to NGC 7469 and IRAS 13120 ($\sim0.8$).

The CN (3-2)/CN (2-1) ratio varies in all the galaxies in our sample (superlinear trends in Table \ref{tab:measured_ratios} and Figure \ref{fig:HCN_and_CN_galax_comps}). The slopes of the CN (3-2)/CN (2-1) ratio in NGC 3256 and IRAS 13120 are smaller ($1.26\pm0.05$ and $1.11\pm0.04$, respectively) than the strong superlinear trend seen in NGC 7469 ($1.55\pm0.04$). We find a higher CN (3-2)/CN (2-1) ratio of $0.45\pm0.06$ in the nucleus of NGC 7469 compared to the non-nuclear value of $0.30\pm0.04$. The nuclear CN (3-2)/CN (2-1) ratio in NGC 7469 more closely resembles the ratio in IRAS 13120 ($\sim0.5$) than NGC 3256 ($\sim0.2$). Fitting the pixels from all galaxies, the CN (3-2)/CN (2-1) ratio has a superlinear slope of $1.33\pm0.02$ (Figure \ref{fig:HCN_and_CN_galax_comps}d). The superlinear trend in this CN (3-2)/CN (2-1) ratio is stronger than the trend in the HCN (4-3)/HCN (3-2) ratio, with values of $1.33\pm0.02$ and $1.13\pm0.03$, respectively.

\subsection{Subtle variations in CN/HCN intensity ratios}
\label{sec:HCN_CN_ratios_results}


We find a higher CN (1-0)/HCN (1-0) intensity ratio of $1.7\pm0.1$ in the northern nucleus of NGC 3256, compared to the non-nuclear value of $1.14\pm0.09$. The global CN (1-0)/HCN (1-0) ratio in IRAS 13120 is also $1.7\pm0.1$. The CN (1-0)/HCN (1-0) ratios in IRAS 13120 and in the northern nucleus of NGC 3256 are significantly higher than the ratio of $\sim1.1$ in NGC 7469. The southern nucleus and non-nuclear regions of NGC 3256 also have smaller ratios of $\sim1.1$.

The variations in the CN (1-0)/HCN (1-0) ratio in NGC 3256 result in a superlinear slope of $1.49\pm0.06$ (Figure \ref{fig:HCN_CN_galax_comps}a). Despite the lack of obvious variations in the different regions of NGC 7469 and IRAS 13120, the slopes indicate subtle variations in the CN (1-0)/HCN (1-0) ratio with values of $1.26\pm0.05$ for NGC 7469 and $0.83\pm0.05$ for IRAS 13120.

For the CN (3-2)/HCN (3-2) ratio, we measure the same ratio of $0.8\pm0.1$ for all regions in NGC 7469 and IRAS 13120.  The CN (3-2)/HCN (3-2) ratio in NGC 3256 has lower values between $0.35-0.55$ (Figure \ref{fig:HCN_CN_galax_comps}b). The lower CN (3-2)/HCN (3-2) ratio in NGC 3256 occur primarily toward the southern half of the galaxy (Figure \ref{fig:CN_HCN_spatial_maps}c). In NGC 3256, the highest CN (3-2)/HCN (3-2) ratio of $0.55\pm0.08$ occurs in the northern nucleus.

\section{Potential Physical Drivers for the Observed Ratio Variations}
\label{sec:discussion}

\subsection{Driving the excitation of HCN}
\label{sec:HCN_excitation}

Molecular excitation is dependent on the temperature, density, and optical depth of the molecular gas, as well as the presence of an external radiation source, such as a PDR and/or an XDR \citep{Boger2005, Meijerink2005, Meijerink2007}. Additionally, molecular excitation can be affected by shocks, which will increase the temperature in the shocked regions of the gas \citep{Martin2015}. The J = 1-0, 3-2 and 4-3 rotational transitions of HCN have energies of $4.25$ K, $25.52$ K and $42.53$ K above the ground state, respectively\footnote{https://home.strw.leidenuniv.nl/$\sim$moldata/}. HCN excitation to the J = 3-2 and 4-3 transitions thus requires much warmer, dense gas conditions than the J = 1-0 transition. An increase in HCN excitation and intensity ratios is expected in regions surrounding AGN sources \citep{Boger2005, Meijerink2005, Meijerink2007, Izumi2013, Izumi2016}. In addition, \citet{Saito2018} find a positive correlation between HCN excitation and SFR surface density when excluding AGN contributions from their analysis. Therefore, we expect to find increased HCN intensity ratios in regions with starburst and/or AGN activity. All galaxies in our sample host some combination of AGN and starburst activity that we often cannot separate at our resolutions. 

\subsubsection{Increased nuclear HCN (3-2)/HCN (1-0) ratio in NGC 7469}
\label{sec:HCN_discuss_7469}

The nuclear region of NGC 7469 has an HCN (3-2)/HCN (1-0) ratio of $\sim0.5$. There is a circumnuclear disk of cold molecular gas surrounding the AGN of this galaxy and forming stars in the inner $\sim100$ pc region \citep{Davies2004, Izumi2015}. Our observations of NGC 7469 have a resolution of $\sim300$ pc and thus blend the influence of both AGN and starburst activity in the circumnuclear disk. This blending makes it challenging to disentangle the effect of the AGN and starburst on the molecular gas, as both will help to increase the HCN (3-2)/HCN (1-0) ratio.

There is evidence of an XDR that is $<50$ pc in radius around the AGN in NGC 7469 \citep{Izumi2015, Izumi2020}. This XDR will influence the higher HCN excitation we find in the nuclear region. \citet{Izumi2013, Izumi2016} conclude that there will be high intensities in the HCN sub-mm lines due to IR-pumping in the presence of XDRs. This IR-pumping is a result of X-ray radiation heating dust grains in the vicinity of the XDR and producing strong IR radiation that affects the HCN chemistry. Our future non-LTE analysis with accurate modelling of gas temperatures and densities will help interpret the HCN (4-3)/HCN (3-2) and HCN (3-2)/HCN (1-0) ratios we find in the nucleus of NGC 7469.

Another plausible scenario for the higher HCN line ratios we find around the AGN in NGC 7469 could be shocks from the AGN outflow increasing the temperature of the gas. Shocked gas will have increased temperatures that favour the warm gas chemistry required to produce HCN \citep{Harada2013, Martin2015}. \citet{Martin2015} find enhanced HCN abundances around the Seyfert 1 AGN in NGC 1097 due to the shocked material produced by the outflows. The peak HCN intensity in their observations is found $\sim200$ pc from the galaxy centre, where the X-ray radiation is weaker and the shocks at the base of the outflow are prominent. At our resolution, we cannot distinguish between the effects of the central XDR in NGC 7469 and any shocks in the outflowing gas that could also increase the gas temperature and affect the HCN excitation in this galaxy.

\subsubsection{Higher HCN (3-2)/HCN (1-0) ratios in IRAS 13120 than NGC 3256}
\label{sec:HCN_discuss_3256_13120}

The HCN (3-2)/HCN (1-0) ratio should increase with increasing starburst contribution due to the positive correlation between HCN excitation and SFR surface density \citep{Saito2018}. \citet{Wilson2019} find ranges of $\Sigma_{\text{SFR}}$ between 10-100 M$_\odot$ yr$^{-1}$ kpc$^{-1}$ for IRAS 13120 compared to 1-25 M$_\odot$ yr$^{-1}$ kpc$^{-1}$ for NGC 3256. \citet{Teng2015} find the 2-10 keV absorption-corrected luminosity of IRAS 13120 is $1.25\times10^{43}$ erg s$^{-1}$ and thermal and non-thermal components of the 0.5-2 keV luminosity are consistent with a star formation rate of $\sim170$ M$_\odot$ yr$^{-1}$. The lower $\Sigma_{\text{SFR}}$ in NGC 3256 could lead to less HCN excitation \citep{Saito2018}. IRAS 13120 has the highest HCN (3-2)/HCN (1-0) ratio ($\sim0.8$), while the northern nucleus of NGC 3256 is lower ($\sim0.4$). We note that the different resolutions in our observations of NGC 3256 ($\sim470$ pc) and IRAS 13120 ($\sim700$ pc) mean that we are comparing different spatial scales in the starburst regions of these two galaxies, which might lead to a difference in the HCN (3-2)/HCN (1-0) ratios.

There is no difference in the HCN (3-2)/HCN (1-0) ratio for IRAS 13120 between the global ratio and the continuum peak. Our resolution of $\sim700$ pc is insufficient to isolate the nuclear region and so the line ratio will have contributions from both the starburst and AGN components in this galaxy. \citet{Veilleux2013} find that IRAS 13120 has a fast, wide-angle outflow seen in the FIR OH line and suggest an AGN contribution to $L_{\text{IR}}$ up to $33.4$\%. This outflow could impact the excitation of HCN through the presence of shocks that could increase the temperature of the gas \citep{Martin2015}.

The southern nucleus of NGC 3256 contains an embedded AGN. Previous work has described this AGN as being inactive or dormant \citep{Sakamoto2014}. The strength of the X-ray emission from this AGN would likely be lower than the other AGN in our sample and we would not expect as significant an increase in the HCN excitation due to X-rays here. \citet{Alonso2012} place a 5\% upper limit on the contribution from any AGN to the total bolometric luminosity of NGC 3256. Both nuclei in NGC 3256 have been detected in X-rays with Chandra by \citet{Lira2002}, although, the authors found no evidence for an AGN in either nucleus. \citet{Ohyama2015} used \textit{Spitzer} data and SED fitting to starburst and AGN templates to suggest there is an AGN in the southern nucleus with $L_{\text{8-1000 \textmu m}} \sim10^{9.7}L_{\odot}$. The northern nucleus is more likely to host an extreme starburst than an AGN \citep{Neff2003}. These limitations on the AGN in NGC 3256 indicate that the HCN excitation conditions will be almost completely driven by starburst effects on the molecular gas. Any contribution from the AGN in this galaxy to HCN excitation would more likely come from shocks in the outflows from the AGN (e.g. \citealt{Aalto2012, Martin2015}).

With higher resolution images and the ability to resolve individual XDRs, we would be able to compare individual ratios based on the relative strengths of the X-rays and outflows created by the AGN.

\subsection{Driving the excitation of CN}
\label{sec:CN_excitation}

As with HCN, the excitation conditions of CN will depend on the temperature, density, optical depth, and presence of external radiation fields in the molecular gas. The CN N = 1-0, 2-1, and 3-2 lines have energies of $5.4$ K, $16.3$ K, and $32.6$ K above ground, respectively. As with HCN, gas traced by higher CN line excitation will be both warmer and/or denser than the N = 1-0 line.

\subsubsection{Superlinear trend in CN (3-2)/CN (2-1)}
\label{sec:CN_discuss_32_10}

The superlinear trend we see in the CN (3-2)/CN (2-1) intensity ratio indicates that the different global ratios between our galaxies are correlated with increased intensity of line emission. We observe higher intensities on average in IRAS 13120 than in NGC 3256 and NGC 7469 for both CN lines. These higher ratios could indicate molecular gas that is more dense with higher temperatures, leading to higher CN excitation. Additionally, because IRAS 13120 is a ULIRG with a high SFR surface density, we could expect a higher external UV field interacting with the molecular gas and leading to an increase in CN abundance by activating the CN formation pathways (i.e. HCN photodissociation and ionization of carbon; \citealt{Boger2005}).

In NGC 7469, the CN (3-2)/CN (2-1) ratio appears to be lowest toward the edges of the map, with mid-range values in the gas surrounding the nucleus, and the highest ratio in the nuclear pixels. We find that the CN (3-2)/CN (2-1) ratio increases from the non-nuclear value of $0.30\pm0.04$ to a value of $0.45\pm0.06$ in the nucleus. The increased CN excitation in this galaxy can thus be attributed to the conditions near the central AGN and surrounding starburst ring. It is apparent that the environment in the nucleus of NGC 7469 has a significant impact on the higher CN excitation. X-rays from the AGN will enhance the bulk temperature of the gas, while PDRs will have higher surface temperatures. In this way, the heating of the gas in the centre of NGC 7469 could be reflected in the CN excitation. We note that it is unclear how far the X-rays will reach, however, and that at our resolution of $\sim300$ pc in this galaxy we do not resolve the XDR to localize the extent of AGN gas heating.

While we do not find any localized enhancement in the CN (3-2)/CN (2-1) ratio in IRAS 13120, the CN (3-2) emission is high throughout the galaxy. We find higher CN (3-2)/CN (2-1) ratios ($>0.5$) in both the global and continuum peak measurements of IRAS 13120 than in any regions of NGC 3256 ($<0.3$) or the non-nuclear region of NGC 7469 ($\sim0.3$). The nuclear region of NGC 7469 has a CN (3-2)/CN (2-1) ratio of $0.45$, that is similar to the global ratio in IRAS 13120. Our ability to resolve different contributions to the excitation to CN (3-2) in IRAS 13120 is limited, while we are better able to localize the effect of the AGN on this ratio in NGC 7469. In any case, both the starburst and AGN contributions in IRAS 13120 will help to increase the excitation of CN to higher values than we find in NGC 3256 and the non-nuclear region of NGC 7469.

We also find a local increase in the CN (3-2)/CN (2-1) ratio in the northern nucleus of NGC 3256 ($\sim0.27$) compared to the southern nucleus ($\sim0.22$) and non-nuclear pixels ($\sim0.21$). However, these ratios are not significantly different once uncertainties on the measured values are taken into account. The higher CN (3-2)/CN (2-1) ratio in the northern nucleus of NGC 3256 could be caused by the more widespread starburst in the northern nucleus. At our resolution ($\sim470$ pc), we do not completely resolve the edge-on southern nucleus and therefore will have lower excitation gas from non-nuclear regions mixed into the beam. \citet{Sakamoto2014} also observed enhanced CN emission associated with the molecular outflows in the northern nucleus of NGC 3256. The CN emission could thus be enhanced as a result of outflow chemistry that is influenced by far-UV emission from massive stars that are formed in the outflow (e.g. \citealt{Cicone2020}). \citet{Cicone2020} note, however, that CN enhancement in the outflow of NGC 3256 is weaker than in Mrk 231.

\subsubsection{The CN (2-1)/CN (1-0) intensity ratio}
\label{sec:CN_discuss_offset_3256}

The CN (2-1)/CN (1-0) ratio has a consistent slope ($\sim0.9$) and ratio ($\sim0.8$) in NGC 7469 and IRAS 13120. This result suggests that the conditions in these two galaxies produce similar excitation conditions for CN. Radiation in the nuclear region of NGC 7469 is dominated by X-rays from the AGN. However, at our resolution of $\sim304$ pc, we will include more than just the X-ray effects on the circumnuclear disk of molecular gas contained within the central beam. The 2-10 keV luminosity of the type 1 Seyfert nucleus is $L_{\text{2-10 keV}} = 1.5\times10^{43}$ erg s$^{-1}$ \citep{Liu2014}. The AGN accounts for $30-40$\% of the IR luminosity in NGC 7469 \citep{Alonso2012}. IRAS 13120 has a Compton thick AGN that accounts for $17-33.4$\% of the total IR luminosity \citep{Iwasawa2011, Teng2015, Veilleux2013}. \citet{Teng2015} find the 2-10 keV absorption-corrected luminosity of IRAS 13120 is $1.25\times10^{43}$ erg s$^{-1}$, which is comparable to that in NGC 7469. At our resolution, we are unable to distinguish between effects from the AGN and/or starburst activity in IRAS 13120. The presence of both UV and X-ray radiation fields should increase the CN abundance \citep{Boger2005, Meijerink2005, Meijerink2007}. Our results indicate that the different radiation fields in NGC 7469 and IRAS 13120 seem to affect the molecular gas in a similar way. 

The offset to lower CN (2-1)/CN (1-0) ratios on the whole in NGC 3256 is interesting. The CN (1-0) data for NGC 3256 is our only data set from ALMA Cycle 0. As such, the flux measurements and intensity ratios using this line in this galaxy have the highest calibration uncertainties of our sample. Given the similarity of the slopes in the CN (2-1)/CN (1-0) ratio of all three galaxies, the offset in the ratio in NGC 3256 could be due to large observational uncertainties. The difference in the CN (2-1)/CN (1-0) ratio in NGC 3256 of $\sim0.45$ compared to the ratios in NGC 7469 and IRAS 13120 of $\sim0.8$ (Figure \ref{fig:HCN_and_CN_galax_comps}) should be treated with caution until this is confirmed with follow-up observations. If real, the offset to lower ratios could be due to different excitation conditions in the dense gas of this galaxy. The small contribution from any AGN in NGC 3256 ($<5$\%; \citealt{Alonso2012}) could perhaps contribute to the lower ($<0.5$) CN (2-1)/CN (1-0) ratio compared to those in NGC 7469 and IRAS 13120 (both $>0.75$). The X-ray luminosity in NGC 3256 is $L_{\text{2-10 keV}} = 1.5\times10^{40}$ erg s$^{-1}$ \citep{Ohyama2015}, three orders of magnitude smaller than both NGC 7469 and IRAS 13120. The models by \citet{Meijerink2007} indicate that CN abundance will be more strongly affected in the presence of XDRs than PDRs, suggesting that the higher CN intensity ratios we find in IRAS 13120 and NGC 7469 could be a result of the influence of the AGN. If both the UV and X-ray fields created in NGC 3256 are weaker than those found in IRAS 13120 or NGC 7469, the CN (1-0) emission could dominate over the CN (2-1) as conditions are not favourable for CN to be excited to higher levels. Future non-LTE modelling to estimate the opacities, temperatures, and densities in these galaxies will allow us to better explain the offset we see in the CN (2-1)/CN (1-0) ratio in NGC 3256.

\subsection{The CN/HCN intensity ratio as a probe of radiation field}
\label{sec:CN_HCN_discuss}

The HCN J = 1-0 and CN N = 1-0 transitions have energies of $\sim4.25$ K and $\sim5.4$ K, respectively, suggesting they trace molecular gas of similar temperatures and densities. In fact, recent work using ALMA observations has found that the global observed CN (1-0)/HCN (1-0) intensity ratio is nearly constant over 3 orders of magnitude change in $\Sigma_{\text{SFR}}$ across a number of different galactic environments, making CN a potentially novel molecule with which to study dense gas (Wilson et al. \textit{in prep}). The average CN (1-0)/HCN (1-0) ratio\footnote{In their analysis, which includes the galaxies NGC 3256, NGC 7469, and IRAS 13120, they consider only the strong CN (1-0) hyperfine line, which accounts for $\sim70\%$ of the total CN (1-0) line emission for optically thin lines \citep{Shirley2015, Meier2015}.} in their sample of 9 galaxies is $0.866\pm0.096$, with a standard deviation of $0.29$. Wilson et al. (\textit{in prep}) use an additional criteria of detecting 93 GHz radio continuum emission in order to compare their CN (1-0)/HCN (1-0) ratios to $\Sigma_{\text{SFR}}$. Since we are not requiring 93 GHz continuum detections, we can extend our analysis to the lower line intensities where we find most of our variations in the CN (1-0)/HCN (1-0) ratios. Including both CN N = 1-0 hyperfine lines, we find ratios of $1.1-1.7$ for the CN (1-0)/HCN (1-0) ratio.

We find variations in the observed intensity ratios between CN and HCN lines, particularly in regions associated with high starburst activity. For example, the CN (1-0)/HCN (1-0) ratio in IRAS 13120 has a value of $1.7\pm0.1$. IRAS 13120 is a ULIRG with a higher $\Sigma_{\text{SFR}}$ than both NGC 3256 and NGC 7469. The increased star formation activity in this galaxy should be indicative of an increase in UV radiation due to young, massive star formation, which could indicate that the UV radiation more strongly affects the gas properties. The northern nucleus of NGC 3256 has the same CN (1-0)/HCN (1-0) ratio of $1.7\pm0.1$ and is also a starburst with a high UV field \citep{Sakamoto2014}. These results will be explored further using non-LTE analysis to calculate the CN/HCN abundance ratio in the presence of UV fields using a PDR model in a future paper.

The CN (1-0)/HCN (1-0) ratio in the southern nucleus of NGC 3256 ($\sim1.1$) is more comparable with that of NGC 7469 ($\sim1.1$). In the southern nucleus of NGC 3256, the AGN and disk are more compact and therefore the sphere of influence of any enhancement from a UV field due to star formation or X-rays from the AGN is relatively small compared to the extent of the starburst in the northern nucleus. The exact impact of X-ray versus UV radiation fields on CN and HCN excitation is beyond the scope of this work.

We find a relatively constant CN (1-0)/HCN (1-0) intensity ratio across NGC 7469, suggesting that the presence of an AGN and XDR does not affect the ratio as significantly as the starbursts do in IRAS 13120 and the northern nucleus of NGC 3256. Non-LTE modelling to determine the CN and HCN abundance ratio is needed to determine if the CN/HCN abundance ratio is also constant. \citet{Izumi2015} suggest that high-temperature gas-phase chemistry in NGC 7469 could significantly enhance the abundance of HCN through an activated formation pathway from CN when T $>300$ K (e.g. \citealt{Harada2010}).

Throughout this paper, we have described the CN/HCN intensity ratios from observations of multiple CN and HCN lines in three galaxies. We find variations in the CN (1-0)/HCN (1-0) intensity ratio in NGC 3256, from values of $\sim1$ in the non-nuclear pixels to values of $\sim1.7$ in the northern nucleus. A simple LTE calculation suggests that the CN and HCN column densities should be in the range of $10^{13}-10^{17}$ cm$^{-2}$. We used this column density range to perform a simple non-LTE calculation with the online version of RADEX\footnote{https://home.strw.leidenuniv.nl/~moldata/radex.html} \citep{vanderTak2007}. Using a gas temperature of $40$ K, an H$_2$ number density of $10^7$ cm$^{-3}$, and a line width of $10$ km s$^{-1}$, we reproduce CN (1-0)/HCN (1-0) $\sim1$ in the non-nuclear pixels of NGC 3256 with a column density ratio of N\textsubscript{CN}/N\textsubscript{HCN} $\sim10$. The intensity ratio of CN (1-0)/HCN (1-0) $\sim1.7$ in the northern nucleus of NGC 3256 is reproduced with N\textsubscript{CN}/N\textsubscript{HCN} $\sim24$. Further non-LTE analysis will be performed in future work, with more analysis on the specific gas properties and optical depths in NGC 3256, NGC 7469, and IRAS 13120 that can produce our observed CN and HCN intensity ratios. In this way, we will be able to compare our data to the expectation that the CN/HCN abundance ratio should increase in the presence of a starburst or increased radiation field (c.f. \citealt{Fuente1993, Fuente1995, Aalto2002, Boger2005, Meijerink2005}.

\section{Conclusions}
\label{sec:conclusions}

In our observational study of CN and HCN lines in a sample of three U/LIRG galaxies, we find subtle variations in line intensity ratios within and among galaxies. These variations can be localized to regions of starburst and/or AGN activity. Both CN and HCN excitation are enhanced in the higher intensity regions.

Our main conclusions are summarized as follows:
\begin{enumerate}
    \item We find variations in the HCN (3-2)/HCN (1-0) ratio in NGC 3256 and NGC 7469. Both galaxies have superlinear trends with slopes $>1.5$. The global and non-nuclear HCN (3-2)/HCN (1-0) and HCN (4-3)/HCN (1-0) ratios are the same ($\sim0.3$) in the two galaxies, which suggests similar excitation conditions. The enhanced HCN excitation is seen in both nuclei of NGC 3256 and in the nucleus of NGC 7469. The HCN (3-2)/HCN (1-0) ratio in NGC 7469 shows the largest difference between non-nuclear ($0.33\pm0.04$) and nuclear ($0.52\pm0.06$) regions. We attribute the regions of enhanced HCN excitation in NGC 7469 to the central AGN, where both an XDR and shocks in outflows can heat the gas and increase HCN excitation. The HCN (3-2)/HCN (1-0) ratios are increased relative to the disk in both nuclei of NGC 3256 due to both starburst and AGN effects, although to a lesser extent than in the nucleus of NGC 7469.
    \item The HCN (3-2)/HCN (1-0) ratio in IRAS 13120 is higher ($\sim0.8$) than both NGC 3256 ($\sim0.27-0.39$) and NGC 7469 ($\sim0.33-0.52$). We find little spatial variation in the HCN excitation of IRAS 13120. We attribute the higher HCN (3-2)/HCN (1-0) ratios to the higher star formation rate surface density and contribution of the Compton-thick AGN to the total IR luminosity.
    \item We find relatively constant CN (2-1)/CN (1-0) ratios and subtle variations in the CN (3-2)/CN (2-1) ratio in our galaxy sample. We find an offset to lower ratios of $\sim0.45$ in the CN (2-1)/CN (1-0) ratio in NGC 3256 compared to $\sim0.8$ in NGC 7469 and IRAS 13120. We note that this offset could arise from the difference in the fraction of AGN contribution to the total IR luminosity in NGC 3256 ($<5$\%) compared to NGC 7469 ($32-40$\%) and IRAS 13120 ($17-33.4$\%). Further non-LTE modelling analysis may help explain the origin of this offset.
    \item The CN (1-0)/HCN (1-0) ratio is higher ($1.7$) in the northern nucleus of NGC 3256 than in the rest of the galaxy. We find a similarly high CN (1-0)/HCN (1-0) ratio in the starburst galaxy IRAS 13120. The intense starburst activity in the regions with large CN (1-0)/HCN (1-0) ratios suggests the enhanced ratio may be due to increased exposure to UV radiation fields. We note that opacity effects could potentially be important and determining the CN/HCN abundance ratios is the next step in this analysis. The CN (1-0)/HCN (1-0) ratio in NGC 7469 is $\sim1.1$ and is relatively constant across the entire galaxy. NGC 7469 is dominated by a nuclear AGN, which does not appear to affect the CN/HCN ratio as significantly as the starbursts in the other galaxies.
\end{enumerate}

The more detailed examination of CN and HCN conducted in this work helps confirm the conclusion of Wilson et al. (\textit{in prep.}) that CN is a viable tracer of dense gas in galaxies. However, it is clear that more work is needed to understand the complex relationship between CN and HCN in dense gas exposed to strong UV and X-ray radiation fields. Future non-LTE radiative transfer modelling of the CN and HCN lines in this galaxy sample will allow us to determine physical properties such as gas temperature, density, opacity, and molecular abundance. Using the molecular abundances, we will be able to compare our results to models predicting CN/HCN abundance ratios and further explore the effects of radiation fields on molecular gas properties.

\section*{Acknowledgements}
We thank the anonymous referee for detailed comments that significantly improved the content of this paper. We would like to thank Kazushi Sakamoto for his insightful comments and important feedback on this work. We also thank Nanase Harada for her collaboration and contribution of ALMA data to this work. This paper makes use of the following ALMA data: \\
ADS/JAO.ALMA\#2011.0.00525.S, ADS/JAO.ALMA\#2012.1.00034.S, ADS/JAO.ALMA\#2012.1.00165.S, ADS/JAO.ALMA\#2013.1.00218.S, ADS/JAO.ALMA\#2013.1.00379.S, ADS/JAO.ALMA\#2015.1.00102.S, ADS/JAO.ALMA\#2015.1.00287.S, ADS/JAO.ALMA\#2015.1.00412.S, ADS/JAO.ALMA\#2015.1.00993.S, ADS/JAO.ALMA\#2016.1.00777.S, ADS/JAO.ALMA\#2018.1.00493.S. ALMA is a partnership of ESO (representing its member states), NSF (USA), and NINS (Japan), together with NRC (Canada), MOST and ASIAA (Taiwan), and KASI (Republic of Korea), in cooperation with the Republic of Chile. The Joint ALMA Observatory is operated by ESO, AUI/NRAO, and NAOJ. The National Radio Astronomy Observatory is a facility of the National Science Foundation operated under cooperative agreement by Associated Universities, Inc.

This research has made use of the NASA/IPAC Extragalactic Database (NED), which is funded by the National Aeronautics and Space Administration and operated by the California Institute of Technology, and the SIMBAD database,
operated at CDS, Strasbourg, France \citep{Wenger2000}. This research has also made use of \textsc{Astropy}, a community-developed core \textsc{Python} package for Astronomy \citep{astropy:2013,astropy:2018}. A significant amount of this research has also made use of the \textsc{Matplotlib} \citep{Hunter2007}, \textsc{NumPy} \citep{van2011}, \textsc{Jupyter Notebook} \citep{Kluyver2016} and \textsc{Linmix}\footnote{https://linmix.readthedocs.io/en/latest/src/linmix.html} \citep{Meyers2018} \textsc{Python} packages.

CDW acknowledges financial support from the Canada Council for the Arts through a Killam Research Fellowship. The research of CDW is supported by grants from the Natural Sciences and Engineering Research Council of Canada and the Canada Research Chairs program. BL wishes to acknowledge partial support from a Natural Sciences and Engineering Research Council of Canada Alexander Graham Bell Canada Graduate Scholarship-Master’s (CGS-M) and an Ontario Graduate Scholarship (OGS).

\section*{Data Availability}
This research is based on public data available in the ALMA archive and the exact project IDs can be found in Table \ref{tab:Project_IDs}. The whole dataset and processed data cubes can be obtained from the authors
by request.




\bibliographystyle{mnras}
\bibliography{CN_HCN_biblio} 



\appendix
\section{Moment maps and additional line ratio measures}

\subsection{Integrated intensity maps}
\label{sec:mom_maps}

\begin{table*}
\centering
    \caption{\small{Data reduction imaging properties and line sensitivities.}}
    \begin{tabular}{cccc}
    \hline 
        \small{Imaging property} & \small{NGC 3256} & \small{NGC 7469} & \small{IRAS 13120} \\
        \hline
        Beam size\textsuperscript{\textit{a}} & $2.2''$ & $0.95''$ & $1.1''$ \\
        Velocity resolution (km s$^{-1}$)\textsuperscript{\textit{b}} & $26.43$ & $20.68$ & $20.64$ \\
        Moment 0 map velocity ranges (km s$^{-1}$) & & & \\
        HCN(1-0) & $2584.30-2980.75$ & $4632.72-5087.68$ & $8744.48-9363.68$ \\
        HCN(3-2) & $2584.30-2980.75$ & $4624.08-5079.04$ & $8744.48-9363.68$ \\
        HCN(4-3) & $2584.30-2980.75$ & $4632.72-5087.68$ & $8744.48-9363.68$ \\
        CN(1-0) & $2555.72-3850.79$ & $4503.40-5087.68$ & $8744.48-9363.68$ \\
        CN(2-1) & $1805.72-3021.50$ & $3865.44-4940.80$ & $8003.20-9262.24$ \\
        CN(3-2) & $1905.72-3121.50$ & $3944.76-5082.16$ & $8044.48-9179.68$ \\
        Smoothed cube line sensitivities (mJy beam$^{-1}$)\textsuperscript{\textit{c}} & & & \\
        HCN(1-0) & $0.241$ & $0.280$ & $0.958$ \\
        HCN(3-2) & $1.624$ & $0.478$ & $0.628$\\
        HCN(4-3) & $1.059$ & $0.872$ & $2.166$ \\
        CN(1-0) & $0.635$ & $0.508$ & $1.220$ \\
        CN(2-1) & $1.021$ & $0.277$ & $0.424$ \\
        CN(3-2) & $0.570$ & $0.753$ & $1.035$ \\
    \hline
    \end{tabular}
    \begin{tablenotes}
    \small
        \item \textit{Notes:} \textsuperscript{\textit{a}}The beams were smoothed, rounded, and matched to this resolution for all lines.
        \item \textsuperscript{\textit{b}}The velocity resolution and pixel size were matched between all lines.
        \item \textsuperscript{\textit{c}}The smoothed cube line sensitivities were determined using the same line-free channels as for the dirty cubes. Integrated intensity map $3.5\sigma$ cut-offs used these smoothed cube sensitivities.
    \end{tablenotes}
    \label{tab:imaging_vel_props}
\end{table*}

\begin{figure*}
	\includegraphics[width=0.92\textwidth]{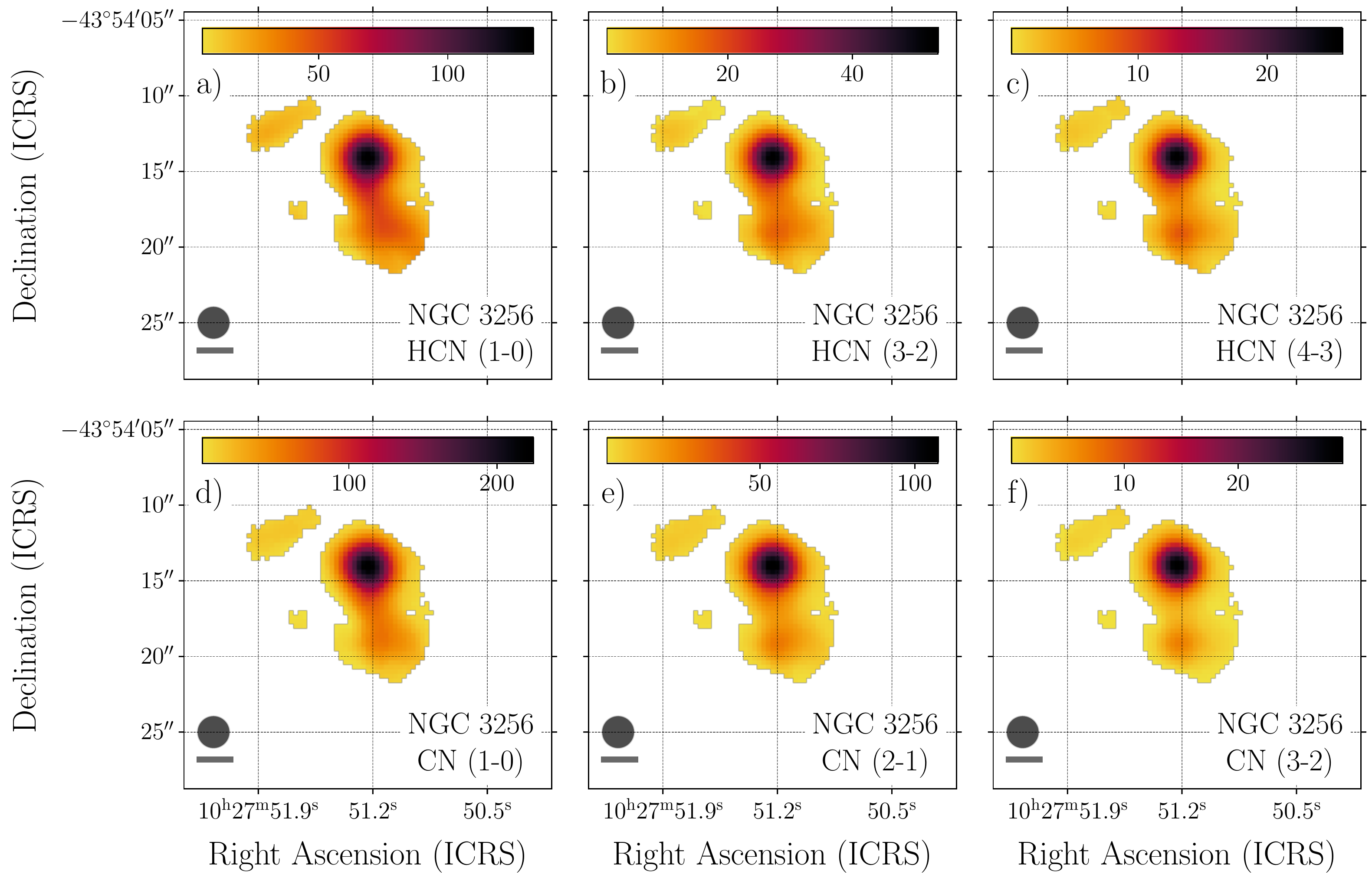}
    \caption{Integrated intensity maps in K km s$^{-1}$ of NGC 3256 for all six lines. In the lower left corner, the circle indicates the beam size of $2.2''$ and the scale bar is set at 500 pc. NGC 3256 clearly has a peak intensity in the northern nucleus. The southern nucleus also shows a local increase in intensity. One spiral arm feature can be seen at our sensitivity in the top left corner of the maps.}
    \label{fig:3256_mom0s}
\end{figure*}

\begin{figure*}
	\includegraphics[width=0.88\textwidth]{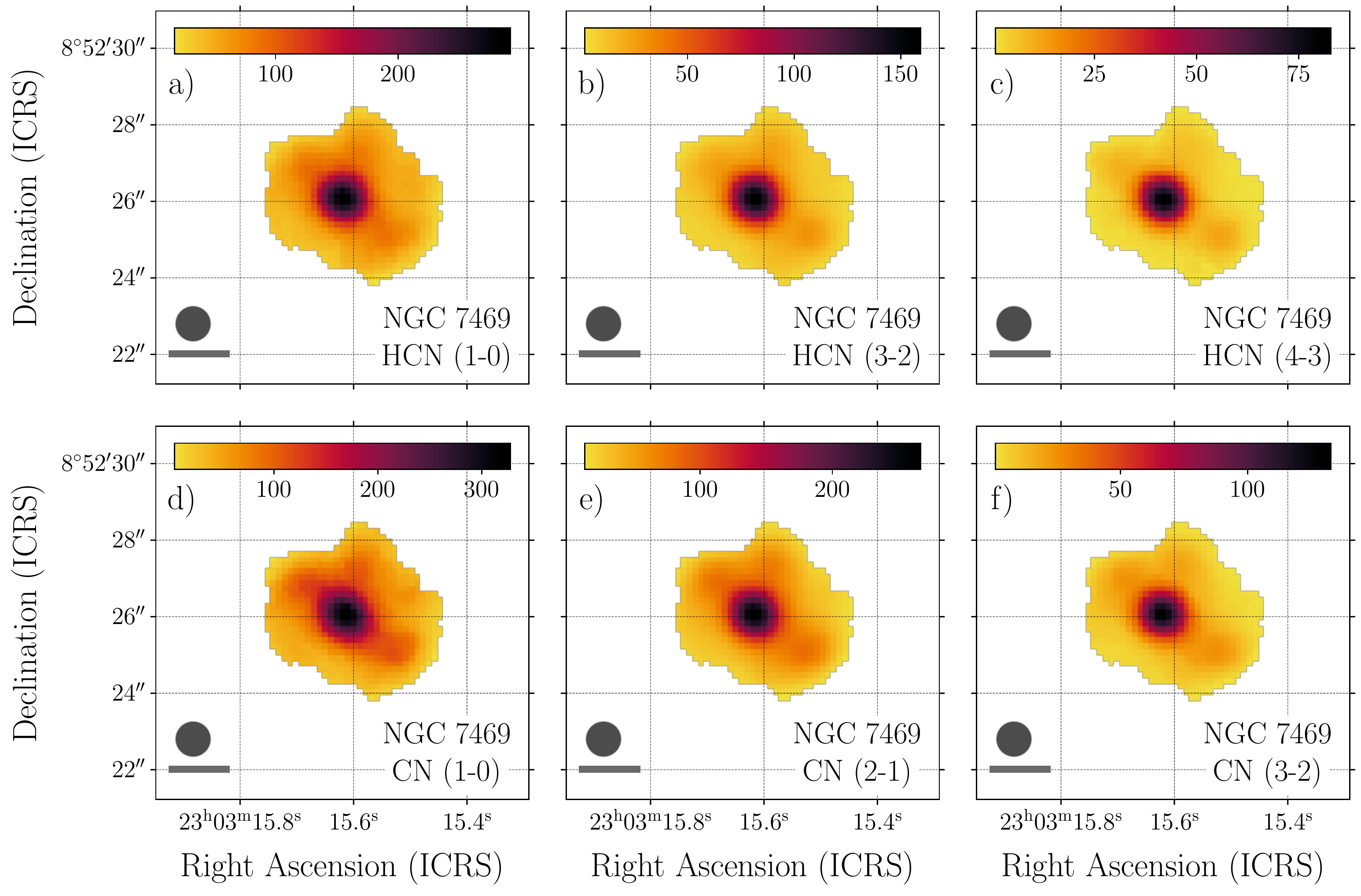}
    \caption{Integrated intensity maps in K km s$^{-1}$ of NGC 7469 for all six lines. In the lower left corner, the circle indicates the beam size of $0.95''$ and the scale bar is set at 500 pc. The peak intensity corresponds to the central nuclear region hosting the AGN. We see some increase in intensity in the region surrounding the nucleus, potentially corresponding to the starburst ring found in this galaxy.}
    \label{fig:7469_mom0s}
\end{figure*}

\begin{figure*}
	\includegraphics[width=0.88\textwidth]{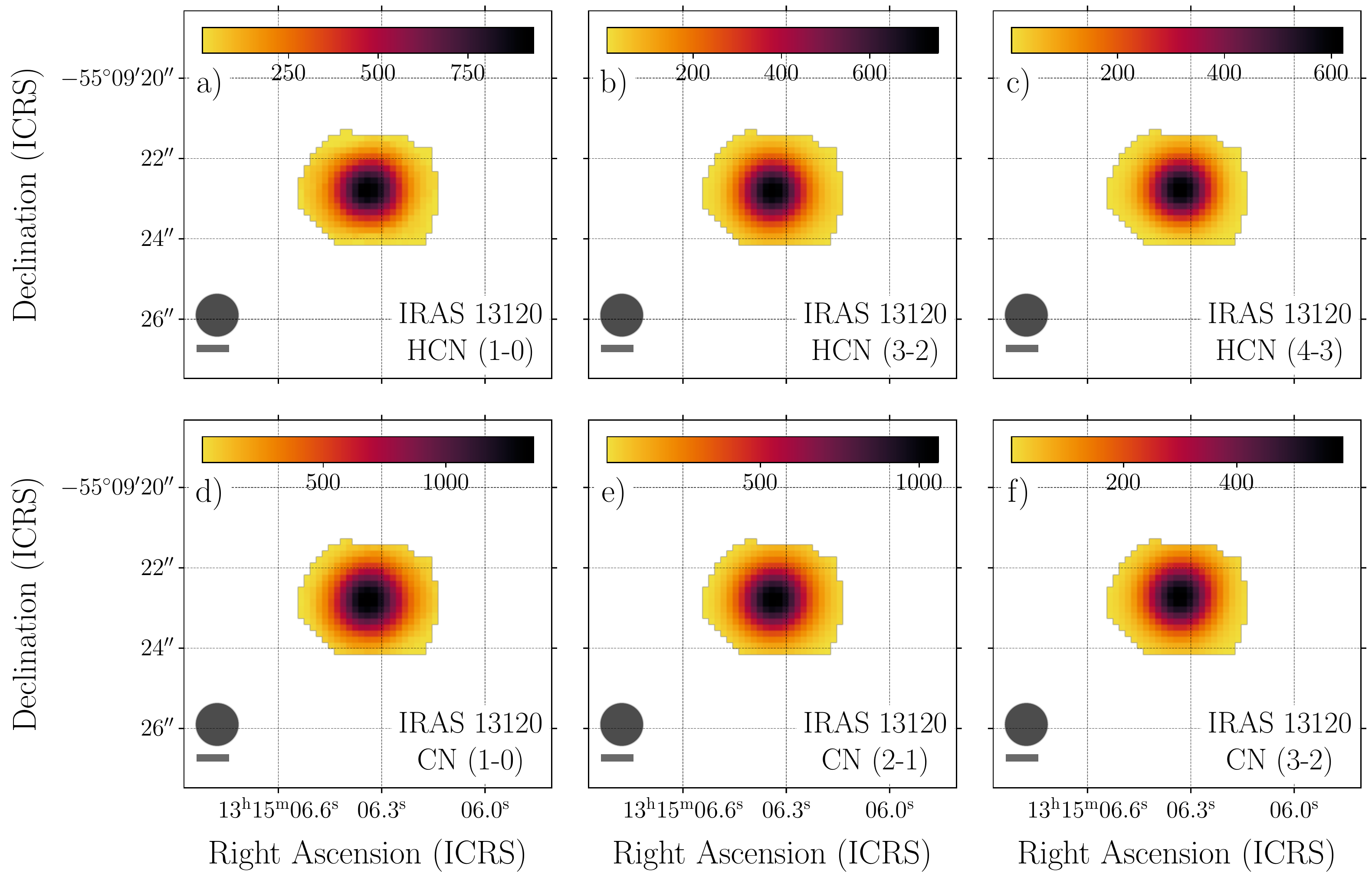}
    \caption{Integrated intensity maps in K km s$^{-1}$ of IRAS 13120 for all six lines. In the lower left corner, the circle indicates the beam size of $1.1''$ and the scale bar is set at 500 pc. The furthest galaxy in our sample, IRAS 13120 has quite a symmetric appearance. The intensities of individual lines at the central peak are much higher than those detected in NGC 3256 and NGC 7469.}
    \label{fig:13120_mom0s}
\end{figure*}

The moment 0 maps of all lines in our three galaxies are shown in Figures \ref{fig:3256_mom0s}, \ref{fig:7469_mom0s}, and \ref{fig:13120_mom0s}. The RMS noise measured from the smoothed cubes and the velocity ranges used are given in Table \ref{tab:imaging_vel_props}.

\subsection{Full measured intensity ratios}
\label{sec:full_ratio_set}

Table \ref{tab:measured_ratios_full} presents the measured CN and HCN intensity ratios in the global, nuclear, and non-nuclear regions of NGC 3256, NGC 7469, and IRAS 13210 for additional pairs of lines beyond those given in Table \ref{tab:measured_ratios}.

\tabcolsep=0.04cm
\begin{table}
\centering
    \caption{\small{Slope and intercept from pixel-by-pixel comparisons.}}
    \begin{tabular}{cccccc}
    \hline
        \small{Ratio} & \small{Params.}\textsuperscript{\text{a}} & \small{NGC 3256} & \small{NGC 7469} & \small{IRAS 13120} & \small{All pixels}\\
        \hline
        \small{$\frac{\text{HCN (3-2)}}{\text{HCN (1-0)}}$} & & & & & \\
        \hline
        & \small{m} & $1.53\pm0.07$ & $1.55\pm0.05$ & $0.84\pm0.06$ & $1.46\pm0.04$ \\
        & \small{b} & $-1.4\pm0.1$ & $-1.45\pm0.08$ & $0.3\pm0.1$ & $-1.26\pm0.07$ \\
        \hline
        \small{$\frac{\text{HCN (4-3)}}{\text{HCN (1-0)}}$} & & & & & \\
        \hline
        & \small{m} & $1.50\pm0.08$ & $2.1\pm0.1$ & $1.19\pm0.07$ & $1.46\pm0.04$ \\
        & \small{b} & $-1.7\pm0.1$ & $-2.9\pm0.2$ & $-0.7\pm0.1$ & $-1.26\pm0.07$ \\
        \hline
        \small{$\frac{\text{HCN (4-3)}}{\text{HCN (3-2)}}$} & & & & \\
        \hline
        & \small{m} & $0.99\pm0.04$ & $1.34\pm0.05$ & $1.3\pm0.1$ & $1.13\pm0.03$ \\
        & \small{b} & $-0.34\pm0.04$ & $-0.91\pm0.06$ & $-0.9\pm0.2$ & $-0.55\pm0.04$ \\
        \hline
        \small{$\frac{\text{CN (2-1)}}{\text{CN (1-0)}}$} & & & & \\
        \hline
        & \small{m} & $0.95\pm0.03$ & $0.93\pm0.03$ & $0.91\pm0.02$ & $1.07\pm0.02$ \\
        & \small{b} & $-0.29\pm0.05$ & $-0.00\pm0.06$ & $0.11\pm0.06$ & $-0.33\pm0.04$ \\
        \hline
        \small{$\frac{\text{CN (3-2)}}{\text{CN (1-0)}}$} & & & & \\
        \hline
        & \small{m} & $1.16\pm0.07$ & $1.42\pm0.07$ & $0.99\pm0.05$ & $1.07\pm0.02$ \\
        & \small{b} & $-1.4\pm0.1$ & $-1.4\pm0.1$ & $-0.4\pm0.1$ & $-0.33\pm0.04$ \\
        \hline
        \small{$\frac{\text{CN (3-2)}}{\text{CN (2-1)}}$} & & & & \\
        \hline
        & \small{m} & $1.26\pm0.05$ & $1.55\pm0.04$ & $1.11\pm0.04$ & $1.33\pm0.02$ \\
        & \small{b} & $-1.05\pm0.06$ & $-1.47\pm0.06$ & $-0.56\pm0.08$ & $-1.11\pm0.03$ \\
        \hline
        \small{$\frac{\text{CN (1-0)}}{\text{HCN (1-0)}}$} & & & & \\
        \hline
        & \small{m} & $1.49\pm0.06$ & $1.26\pm0.05$ & $0.83\pm0.05$ & $1.25\pm0.03$ \\
        & \small{b} & $-0.74\pm0.09$ & $-0.40\pm0.08$ & $0.6\pm0.1$ & $-0.38\pm0.06$ \\
        \hline
        \small{$\frac{\text{CN (2-1)}}{\text{HCN (3-2)}}$} & & & & \\
        \hline
        & \small{m} & $0.89\pm0.05$ & $0.77\pm0.02$ & $0.86\pm0.04$ & $0.91\pm0.02$ \\
        & \small{b} & $0.35\pm0.05$ & $0.70\pm0.03$ & $0.54\pm0.08$ & $0.44\pm0.03$ \\
        \hline
        \small{$\frac{\text{CN (3-2)}}{\text{HCN (3-2)}}$} & & & & \\
        \hline
        & \small{m} & $1.06\pm0.09$ & $1.17\pm0.05$ & $0.94\pm0.06$ & $1.21\pm0.04$ \\
        & \small{b} & $-0.56\pm0.08$ & $-0.35\pm0.06$ & $0.1\pm0.1$ & $-0.52\pm0.05$ \\
        \hline
        \small{$\frac{\text{CN (3-2)}}{\text{HCN (4-3)}}$} & & & & \\
        \hline
        & \small{m} & $1.14\pm0.07$ & $0.85\pm0.03$ & $0.69\pm0.04$ & $1.01\pm0.04$ \\
        & \small{b} & $-0.23\pm0.05$ & $0.46\pm0.03$ & $0.75\pm0.07$ & $0.12\pm0.04$ \\
    \hline
    \end{tabular}
    \label{tab:slopes_full}
    \begin{tablenotes}
        \small
            \item \textit{Notes:} \textsuperscript{\text{a}}Fit parameters are determined from Linmix linear fitting of the form log(y) = m log(x) + b.
    \end{tablenotes}
\end{table}

\tabcolsep=0.2cm
\begin{table*}
\centering
    \caption{\small{Additional CN and HCN measured intensity ratios.}}
    \begin{tabular}{ccccccccccc}
    \hline
        \small{Region} & \small{$\frac{\text{HCN (3-2)}}{\text{HCN (1-0)}}$} & \small{$\frac{\text{HCN (4-3)}}{\text{HCN (1-0)}}$} & \small{$\frac{\text{HCN (4-3)}}{\text{HCN (3-2)}}$} & \small{$\frac{\text{CN (2-1)}}{\text{CN (1-0)}}$} & \small{$\frac{\text{CN (3-2)}}{\text{CN (1-0)}}$} & \small{$\frac{\text{CN (3-2)}}{\text{CN (2-1)}}$} & \small{$\frac{\text{CN (1-0)}}{\text{HCN (1-0)}}$} & \small{$\frac{\text{CN (2-1)}}{\text{HCN (3-2)}}$} & \small{$\frac{\text{CN (3-2)}}{\text{HCN (3-2)}}$} & \small{$\frac{\text{CN (3-2)}}{\text{HCN (4-3)}}$} \\
        \hline
        NGC 3256 & & & & & & & & & & \\
        \hline
        Global & $0.29(3)$ & $0.13(2)$ & $0.46(7)$ & $0.44(5)$ & $0.10(1)$ & $0.22(3)$ & $1.24(9)$ & $1.9(3)$ & $0.42(6)$ & $0.9(1)$ \\
        North nucleus & $0.39(4)$ & $0.18(2)$ & $0.48(7)$ & $0.47(5)$ & $0.13(1)$ & $0.27(4)$ & $1.7(1)$ & $2.1(3)$ & $0.55(8)$ & $1.2(2)$ \\
        South nucleus & $0.33(4)$ & $0.15(2)$ & $0.46(7)$ & $0.44(5)$ & $0.10(1)$ & $0.22(3)$ & $1.12(9)$ & $1.5(2)$ & $0.34(5)$ & $0.8(1)$ \\
        Non-nuclear & $0.27(3)$ & $0.12(1)$ & $0.45(6)$ & $0.43(5)$ & $0.09(1)$ & $0.21(3)$ & $1.14(8)$ & $1.8(3)$ & $0.38(5)$ & $0.8(1)$ \\
        \hline
        NGC 7469 & & & & & & & & & \\
        \hline
        Global & $0.36(4)$ & $0.14(2)$ & $0.40(6)$ & $0.76(9)$ & $0.25(3)$ & $0.32(5)$ & $1.12(8)$ & $2.4(3)$ & $0.8(1)$ & $1.9(3)$ \\
        Nucleus & $0.52(6)$ & $0.25(3)$ & $0.48(7)$ & $0.80(9)$ & $0.36(4)$ & $0.45(6)$ & $1.13(9)$ & $1.7(2)$ & $0.8(1)$ & $1.6(2)$ \\
        Non-nuclear & $0.33(4)$ & $0.12(1)$ & $0.37(5)$ & $0.76(9)$ & $0.22(3)$ & $0.30(4)$ & $1.11(8)$ & $2.6(4)$ & $0.8(1)$ & $2.1(3)$ \\
        \hline
        IRAS 13120 & & & & & & & & & \\
        \hline
        Global & $0.80(9)$ & $0.61(7)$ & $0.8(1)$ & $0.77(9)$ & $0.40(5)$ & $0.53(8)$ & $1.7(1)$ & $1.6(2)$ & $0.8(1)$ & $1.1(2)$ \\
        Continuum peak & $0.79(9)$ & $0.70(8)$ & $0.9(1)$ & $0.80(9)$ & $0.45(5)$ & $0.57(8)$ & $1.5(1)$ & $1.5(2)$ & $0.8(1)$ & $0.9(1)$ \\
    \hline
    \end{tabular}
    \begin{tablenotes}
    \small
        \item \textit{Notes:} The uncertainties presented here are the measurement and calibration uncertainties and given as the uncertainty on the last digit (i.e. $0.29(3) = 0.29\pm0.03$).
        \item Ratios include the 5\% (Band 3) and 10\% (Band 6 and 7) ALMA calibration uncertainties for each line.
    \end{tablenotes}
    \label{tab:measured_ratios_full}
\end{table*}

\subsection{Additional pixel-by-pixel comparisons}
\label{sec:add_pp_comp}

\begin{figure*}
	\includegraphics[width=\textwidth]{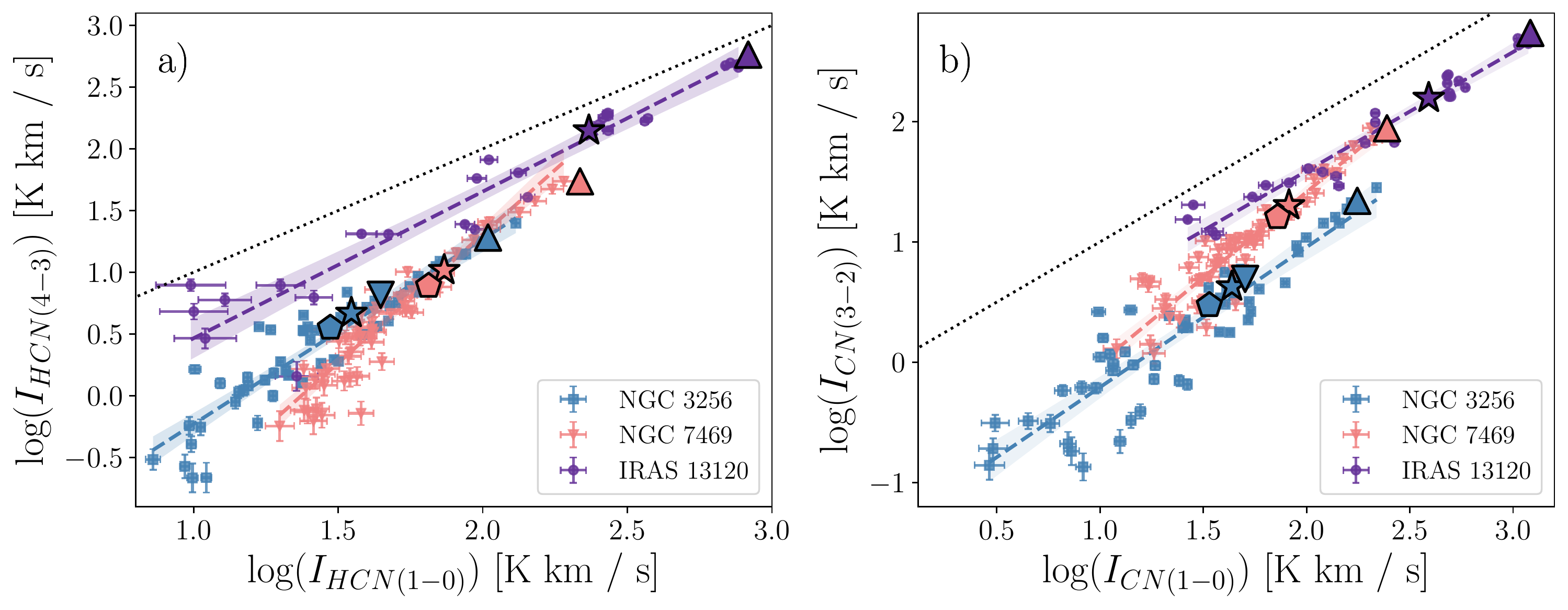}
	\caption{Pixel-by-pixel comparisons of individual line intensities in each galaxy. The higher-J/N transition of the two lines is always plotted on the y-axis. The black dotted line is the one-to-one line. Blue squares correspond to NGC 3256; pink inverted triangles correspond to NGC 7469; purple circles correspond to IRAS 13120. The dashed lines in each colour represent the Linmix fits for each galaxy, with the shaded region representing the 95\% confidence interval of these fits. Large stars show the global ratio in each galaxy and large pentagons show the non-nuclear ratios in NGC 3256 and NGC 7469. Large triangles show ratios for the northern nucleus of NGC 3256, the nucleus of NGC 7649, and the continuum peak in IRAS 13120, while the large inverted triangle shows the ratio for the southern nucleus of NGC 3256. The black dot-dashed lines and shaded regions in (c) and (d) show Linmix fits to all three galaxies.}
   \label{fig:HCN_and_CN_high_lines}
\end{figure*}

\begin{figure*}
	\includegraphics[width=\textwidth]{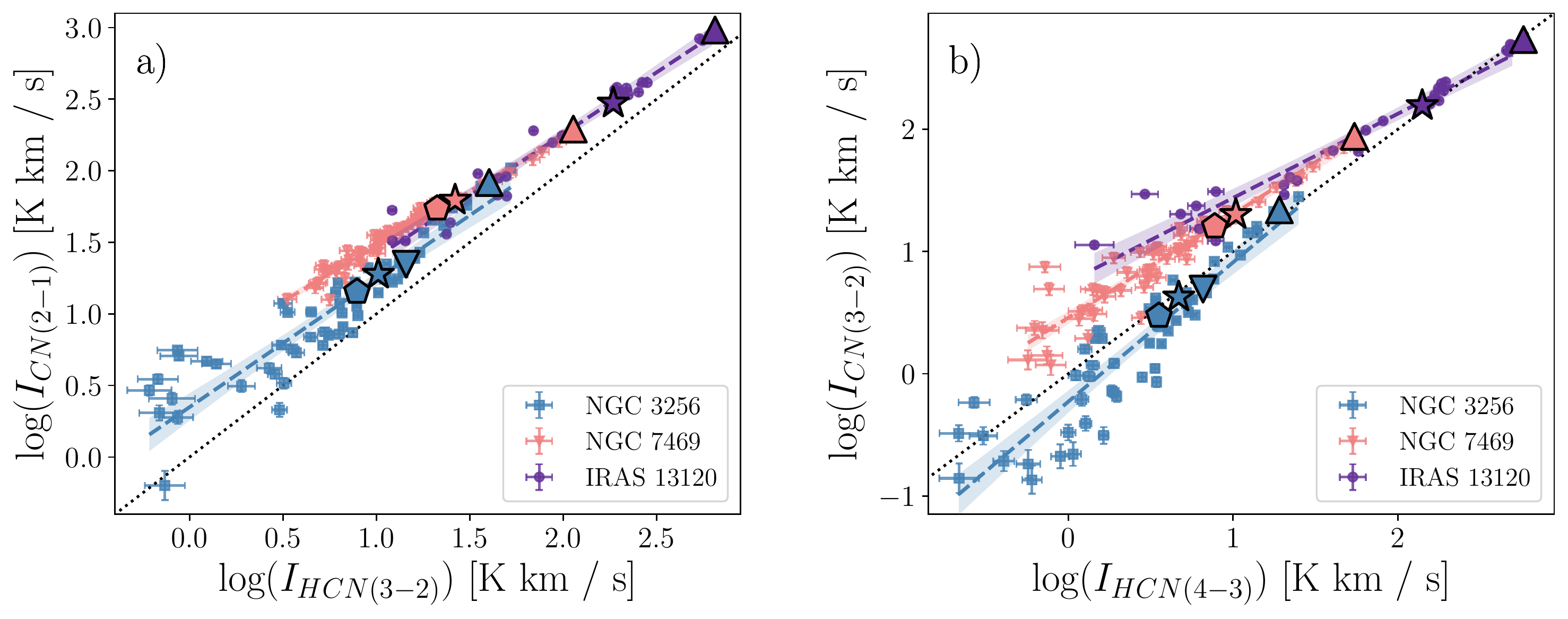}
	\caption{Pixel-by-pixel comparisons for CN compared to HCN for selected transitions. For plot descriptions, refer to the caption of Figure \ref{fig:HCN_and_CN_high_lines}.}
   \label{fig:HCN_CN_off_lines}
\end{figure*}

We present additional pixel-by-pixel correlations between HCN (4-3) and HCN (1-0), CN (3-2) and CN (1-0), CN (2-1) and HCN (3-2), and CN (3-2) and HCN (4-3) using the intensities in the re-sampled moment 0 maps in the scatter plots of Figures \ref{fig:HCN_CN_off_lines} and \ref{fig:HCN_and_CN_high_lines}. In Table \ref{tab:slopes_full}, we present the measured slopes and intercepts from the Linmix linear regression fits to the various line ratios explored in the pixel-by-pixel comparisons. In this table we also include the HCN (4-3)/HCN (1-0), CN (3-2)/CN (1-0), CN (2-1)/HCN (3-2), and CN (3-2)/HCN (4-3) ratios. The linear fits were in the form log(y) = m log(x) + b on to the pixel-by-pixel comparisons of the re-sampled moment 0 maps and we present both m and b values here. Inclination corrections are as described in Section \ref{sec:pixel-by-pixel}. Fits to all pixels from all galaxies for the various line ratios are presented in Table \ref{tab:slopes_full}, as well.

\bsp	
\label{lastpage}
\end{document}